\documentclass[aps,pra,notitlepage]{revtex4-1}
\usepackage{amsmath,graphicx,color}
\usepackage[colorlinks]{hyperref}
\hypersetup{colorlinks,citecolor=cyan,linkcolor=blue,urlcolor=blue}
\newcommand{\be}{\begin{equation}}
\newcommand{\ee}{\end{equation}}
\begin{document}
\title{Fractional Laplacians  in bounded domains:
 Killed, reflected, censored and  taboo  L\'{e}vy flights}
\author{Piotr Garbaczewski and Vladimir Stephanovich}
\affiliation{Institute of Physics,  University  of Opole, 45-052 Opole, Poland }
\begin{abstract}
The  fractional Laplacian $(- \Delta  )^{\alpha /2}$,  $\alpha \in (0,2)$   has many equivalent (albeit formally  different) realizations as a nonlocal   generator of   a family of  $\alpha $-stable   stochastic  processes in $R^n$. On the other hand, if the process is to be restricted to a bounded domain,   there are many inequivalent proposals for what a boundary-data respecting fractional Laplacian   should actually  be. This ambiguity holds true  not only   for each  specific  choice of the  process behavior at the  boundary  (like e.g.   absorbtion, reflection,  conditioning or boundary  taboos), but   extends  as well  to its   particular  technical  implementation (Dirchlet,  Neumann, etc.  problems).  The   inferred  jump-type processes  are inequivalent as well, differing in their  spectral and  statistical  characteristics,    which may strongly   influence the ability of the  formalism (if uncritically adopted) to   provide an unambigous  description of  real geometrically confined physical systems with disorder. Specifically  that refers to their  relaxation properties and the near-equilibrium asymptotic behavior.  In the present paper we focus on  L\'{e}vy flight-induced jump-type processes  which are  constrained to stay forever inside a finite domain.  That  refers to  a concept of taboo processes  (imported  from Brownian to  L\'{e}vy - stable contexts), to so-called censored processes  and  to   reflected  L\'{e}vy flights  whose status still  remains to be unequivocally  settled. As a byproduct of our fractional  spectral analysis, with reference to  Neumann  boundary  conditions,  we discuss disordered semiconducting heterojunctions as the bounded domain problem.
  \end{abstract}
\date{\today}
\maketitle
\section{Motivation}
Brownian motion in a  bounded domain is a classic problem with an ample coverage in the literature, specifically concerning the absorbing (Dirichlet) and reflecting (Neumann) boundary data (for the present purpose we disregard other boundary data choices).  A coverage concerning their physical relevance is enormous as well \cite{schuss,carlsaw}.

 Anticipating further discussion, we  quite inentionally    point out source  papers dealing with  reflected  Brownian motion, \cite{grebenkov} and exposing at  some length   the method of eigenfuction expansions for the  reflected and other  boundary-data problems, \cite{grebenkov1}-\cite{bickel0}, c.f. also \cite{risken}.   The latter method  is as  well    an indispensable tool  in the analysis of spectral properties of fractional Laplacians and related jump-type processes in bounded domains.
  Its direct link with  well developed theory of heat semigroups   for  jump-type processes (mostly these with absorpion/killing)  allows to address   the statistics of exits from  the domain, like e.g.  the  first and mean first exit times,   large time behavior, stationarity  issues, probability of   survival   and its  asymptotic decay, c.f. \cite{lorinczi}-\cite{gar}.
 Compare e.g. also  \cite{gar1,mazzolo} (Brownian case)  and \cite{gaps,kaleta,frank} (L\'{e}vy-stable case), where the role of lowest eigenstates and eigenvalues (thence eigenvalue gaps)   of the motion generator has   appeared to be  vital for the description of decay rates  of killed stochastic processes.  The  spectral data  of motion generators  are relevant for quantifying  long-living processes  in a spatial trap (e.g. bounded domain), eventually with an infinite life-time.

Currently,  the literature devoted to fractional Laplacians and related jump-type processes is extrememly rich, albeit   with no efficient  interplay/communication  between   physics and mathematics-oriented  communities. There is a  definite prevalence of the very active    purely mathematical research  on this  subject matter.  On the other hand,  even a concise listing of various   real-world applications of the fractional calculus in science and engineering is beyond the ramifications  of an introductory section of this paper, see e.g.  for example  \cite{collection} - \cite{laskin}.   Viewpoints of  applied mathematicians can be consulted in \cite{bucur}- \cite{servadei}.

A departure point of the  present work  is  an apparent  incompatibility of  the implementation of  reflecting boundary conditions for L\'{e}vy flights in  the physics-motivated investigations, Refs. \cite{dybiec} - \cite{denisov},  while set against   varied  (inequivalent) proposals available in  the mathematical literature,  Refs.  \cite{bogdan} - \cite{grubb}.

We note that a  general  theory of  censored L\'{e}vy processes has been developed \cite{bogdan}, to handle jump-type processes which are not allowed
to jump-out of an open, sufficiently regular set  $D  \subset R^n $  (eventually closed, under suitable precautions, \cite{guan}).
  Within this theory,  reflected L\'{e}vy-stable processes  have been introduced  and  so-called regional fractional Laplacians  were identified as  generators  of these processes, \cite{refl,guan,warma}.  Nonlocal analogs of the Neumann  boundary data have been    associated with them in suitable ranges of the stability parameter, \cite{guan}.  We point   out that other analogs of nonlocal Neumann data, imposed directly on the fractional Laplacian,  were proposed as well, \cite{dipierro}.

 The existence problem for jump-type processes with an infinite
life-time in a bounded domain, seems to have been left aside in the
physics literature (see however  Refs. \cite{gar1,mazzolo}  in connection with diffusion processes and \cite{zaba,gar}   for a preliminary discussion of the Cauchy process that is trapped  in the interval).  To the contrary, permanently trapped
L\'{e}vy-type processes (diffusion processes like-wise) have their  well established  place in the mathematical literature.

One category of such processes stems from the analysis of the long-time
behavior of the survival probability in the case of absorbing enclosures. One may  actually  single out appropriately  conditioned processes that never leaves the domain once started within.
Another category  can be related  to reflecting boundary data. In contrast to  the   reflected Brownian  motion  this  issue is   conceptually more involved  and  as yet   not free from ambiguities  in  the context of      L\'{e}vy-stable processes.  Thus, both from  physical and applied mathematics points of view,    constructing  well-posed   fractional (L\'{e}vy) transport models  in  bounded domains  and keeping under control  (the degree and physical  relevance of) their  possible inequivalence      is of vital importance.

In the traditional Brownian lore, while giving meaning to the Laplacian in a bounded domain  $D\subset R^n$, denoted tentatively $\Delta _D$, we must account for various admissible boundary data,  that are local i.e.  set at  the boundary $\partial D$  of an  open set $D$.
One may try to  define a fractional power of the Laplacian by importing its  locally defined   boundary data on $\partial D$,   through so-called
 spectral  definition  $(- \Delta _D)^{\alpha /2}$, \cite{bucur}-\cite{what}.
This  operator   is known to be different,  \cite{servadei},  from   the   outcome  of the  procedure  in which one first executes  the fractional power of the Laplacian, and next imposes  the boundary data, as embodied in the notation $(- \Delta )^{\alpha /2}_D$.   In case of absorbing boundaries,  in contrast to $(- \Delta _D)^{\alpha /2}$ where Dirichlet conditions can be  imposed locally at the boundary  $\partial D$  of $D$,
 for   $(- \Delta )^{\alpha /2}_D$   these  data    need to be imposed as exterior  ones i.e. in the  whole  complement $R^n\backslash D$ of $D$.

In passing we  note that it is the nonolocality of  fractional motion generators  (fractional Laplacians)  that is the main source of difficulties if the finite-size domain problems are to be considered.  In the  familiar to physicists lore of Riemann-Liouville fractional  derivatives, defined in the Caputo sense,
   it is known that  the divergence problems arise    near the domain boundaries.
    In the study of  transport properties of magnetically confined plasmas,
   \cite{negrete}, a regularization of the otherwise
 singular fractional    derivative  of a general function,  has been accomplished
  by subtracting the  boundary terms.
  A careful handling of  such  terms  appears to  be vital in the present research, and allows  to    make a clear distinction between e.g. absorbing, censored and reflected processes  and the  corresponding  fractional
     motion generators.

Reflected Brownian motions belong to the bounded domain paradigm, \cite{bickel0,bickel} and  the  general family of censored L\'{e}vy flights (reflected case being included)  likewise.  If one resorts to the
spectral  definition of the fractional Laplacian on $D$, Neumann conditions
can be  imported directly from the Brownian framework  and imposed locally.

This is not the case, if  censored L\'{e}vy processes  and regional Laplacians  enetr the game.   In connection with reflected L\'{e}vy flights, a fairly nontrivial problem is  to deduce a proper  nonlocal analog of the Neumann boundary condition, so that the existence status of the regional generator  (and thence of  the induced  process) can be granted. Even more diffucult issue is to provide a consistent (semi)phenomenological picture of  the reflection mechanism, that should underly (or directly follow from) the   mathematical procedure.  Notwithstanding,   Neumann type problems can be obtained in many ways, depending on the kind of "reinjection" we impose on the outside jumps, \cite{barles}.

The physically appealing reflection mechanism (a limiting infinite well/trapping interval  case of the strongly anharmonic Langevin - L\'{e}vy  evolution, c.f.  Refs.  \cite{dybiec} - \cite{denisov},   cannot be  justified on the  basis of  the  existing  mathematical  theory of reflected L\'{ e}vy processes of Refs. \cite{bogdan,guan,warma}   and  needs  a deepened discussion concerning its  meaning  and range of  validity.     On purely mathematical grounds, the resultant asymptotic probability density function can  be  readily recognized  \cite{gar}  as so-called $\alpha$ - harmonic function of the fractional Laplacian $(-\Delta )^{\alpha /2}_D$. Here we emphasize that  the $\alpha $-harmonic function needs to be defined globally in $R$, albeit    may vanish  beyond $D$, i.e in $R\backslash D$.

A strictly positive  part (restricted exclusively  to $D$) of the pertinent  function, while  normalized on $D$, can be interpreted as  a probability density  and, according to    Ref. \cite{denisov},  sets
a  formal explanation  of  the   "origin of the preferred concentration of flying objects near the boundaries in nonequilibrium systems".

 We note that in  the   interior of $D$ this pdf   rapidly diverges while approaching the boundary,  in the whole parameter range $\alpha \in (0,2)$.  Such probability   accumulation in the vicinity of the boundary   barrier has been reported recently in the analysis  of the fractional Brownian motion with a reflecting wall \cite{wada}, albeit only in the superdiffusive regime $\alpha >1$, while a  probability  depletion close to the  barrier has been a characteristic of the subdiffusive regime $\alpha <1$.

Reflecting barriers were seldom  seriously  addressed by physicists in case of L\'{e}vy -type processes, fractional diffusions and continuous time random walks scenarios.  
On the other hand,  their role in the so-called fractional
 Brownian motion  and general anomalous diffusion problems
   has been analyzed, \cite{neel,metzler,metzler1},  with observations   that are   different  from the previously outlined  ones. As well they remain
  incongruent  with more  mathematically  oriented  (including  computer-assisted)  research on reflected L\'{e}vy flights and fractional diffusion with reflection,  \cite{meer}.

In the present paper, the main body of arguments has its roots in the theory of (nonlocally induced, L\'{e}vy-stable)  Markov  stochastic processes and spectral properties of their (nonlocal as well, fractional) generators.  Hence many interesting resarch streamlines which  refer to varied realisations of anomalous diffusions (processes with memory,those  deriving from the continuous  time random walks (CTRW), standard fractional Brownian motion, etc.)   are left aside. Nonetheless we mention some  source  papers, that investigate relaxation properties in  the fractional    transport that is governed by  generalized Langevin equations (GLE), in particular in a finite domain, \cite{oliveira,kinley,vainstein}  and specifically \cite{wada,metzler2}  in the context of the fractional Brownian motion, where depletion or accretion zones of particles near boundaries have been numerically predicted.

In the latter case \cite{metzler} a clear specification is given of what pragmatically oriented researchers  interpret as a reflection from the boundary (there are different prescriptions that may lead to inequivalent outcomes).  The  reflection  recipe always refers to the trajectory behavior  in the vicinity of the boundary. One needs to state clearly how to execute a reflection  in the Monte Carlo path-wise simulations,  i.e. not to cross the boundary   once a jump of a given length  would    definitely take us away from the trapping enclosure.  Compare e.g. our discussion in section VI and  a related discussion of   reflection conditions  in Refs. \cite{dybiec}-\cite{denisov}. Neither of these papers addresses the reflection boundary conditions for motion generators per se.

\section{Varied   (in)equivalent  faces of  the fractional Laplacian.  }

\subsection{Fractional  Laplacians in $R^n$}
In the present paper, up to    suitable  adjustment  of   dimensional constants,    the free stochastic   evolution  in $R^n$  refers either to the nonnegative motion generator $ - \Delta$  (Brownian motion) or   $(-\Delta )^{\alpha /2}$ with $0<\alpha <2$   (L\'{e}vy-stable motion).  One  should keep in mind that it  is   $-(-\Delta )^{\alpha /2}$ which stands for  a  legitimate  fractional relative  of  the ordinary Laplacian  $\Delta $.

It is known that there are many  formally different definitions of the fractional Laplacian which actually are  equivalent, \cite{kwasnicki}.
For our purposes we  shall reproduce three
equivalent in $R^n$  definitions of the
symmetric L\'{e}vy stable generator, which  nowadays
are  predominantly employed in the literature (we do not directly refer to
the popular notion of a
fractional derivative, although formally one  can  write $(-\Delta )^{\alpha /2} \equiv - \partial ^2/\partial |x|^{\alpha /2}$,  whatever a specific
form of $(-\Delta )^{\alpha /2}$ is chosen).

The   spatially nonlocal fractional Laplacian has an  integral
definition (involving a  suitable function $f(x)$, with $x\in
R^n$)  in terms of the Cauchy principal value  (p.v.), that  is
 valid in space   dimensions $n\geq 1$
 \be
(-\Delta)^{\alpha /2}f(x)=\mathcal{A}_{\alpha,n} \lim\limits_{\varepsilon\to 0^+}
\int\limits_{{R}^n\supset \{|y-x|>\varepsilon\}}
\frac{f(x)-f(y)}{|x-y|^{\alpha +n}}dy.
 \ee
where  $dy \equiv d^ny$ and  the (normalisation) coefficient:
\be
\mathcal{A}_{\alpha,n}=
 \frac{2^{\alpha } \Gamma ({\frac{\alpha + n}{2}})}{\pi ^{n/2}
  |\Gamma (- {\frac{\alpha }{2}})|}  =
  \frac{2^{\alpha } \alpha \Gamma ({\frac{\alpha + n}{2}})}{{\pi ^{n/2}
  \Gamma (1- \alpha /2})}
\ee
Here one needs to employ   $\Gamma (1-s)= -s \Gamma (-s)$ for any $s\in (0,1)$.  The   normalisation coefficient has been adjusted  to secure that  the  integral   definition  stays in conformity   with its Fourier
transformed version. The latter  actually gives  rise to the
widely (sometimes uncritically) used  Fourier multiplier representation  of the fractional Laplacian, \cite{kwasnicki,bucur,abtangelo,what,kwa}:
\be
{\cal{F}} [(- \Delta )^{\alpha /2} f](k) = |k|^{\alpha } {\cal{F}} [f](k).
\ee
We recall again,  that it is $- (-\Delta )^{\alpha /2}$ which is a fractional analog of the Laplacian  $\Delta $.

Another definition, being  quite  popular in the literature  in view of  more explicit
dependence on the ordinary Laplacian, derives directly  from  the standard  Brownian semigroup   $\exp(t\Delta )$ (in passing we note that the L\'{e}vy
 semigroup reads $\exp[-t(- \Delta )^{\alpha /2}]$.     The pertinent semigroup  is explicitly  built into the formula, originally  related to   the  Bochner's  subordination  concept, \cite{kwasnicki,kwa,vondr}:
\be
(-\Delta )^{\alpha /2}f   = {\frac{1}{|\Gamma (-{\frac{\alpha }{2}})|}}\,
 \int_0^{\infty} (e^{t \Delta }f - f) t^{-1-\alpha /2}\, dt .
\ee
Clearly, given an initial datum $f(x)$, we deal here  directly involved a solution  of the  standard  (up to dimensional coefficient)  heat equation $f(x,t) = (e^{t \Delta }f$  in the above  integral formula.

We note, that based on tools from functional analysis   (e.g. the spectral theorem), this definition of the fractional Laplacian
extends to   fractional powers of   more general  non-negative operators, than $(-\Delta )$ proper. This point  will  receive more  attention in below.

{\bf Remark 1:}  While computing the singular integral (1), one needs some care if a decomposition into a sum  of  integrals  is involved.  An alternative definition
\be
(-\Delta)^{\alpha /2}f(x)=  {\frac{\mathcal{A}_{\alpha,n}}{2}} \int _{R^n}
\frac{2f(x) - f(x+y) -f(x-y)}{|y|^{n+\alpha }}\, dy,
\ee
if employed in suitable function spaces,  is by construction free of   singularities  and does not involve  the pertinent decompositions,  \cite{bucur,servadei,abtangelo}.

 \subsection{Fractional Laplacians  in a bounded domain.}

\subsubsection{Restricted  (hypersingular)   fractional Laplacian}

  As mentioned before, a domain restriction to a bounded subset $D \subset R^n$ is hard, if  not  impossible,  to implement via the  Fourier multiplier definition. The reason is an inherent spatial nonlocality of L\'{e}vy-stable generators.

We confine attention to   the    Dirichlet  boundary data in  a bounded domain. Here one begins from the formal  fractional operator  definition in $R^n$   and restricts its action to suitable  functions with  support in $D\subset R^n$.  It is known that  the  standard Dirichlet restriction $f(x)= 0 $  for  all $x\in \partial D$ is insufficient. One needs to impose the  so-called exterior  Dirichlet condition:   $f(x) = 0$ for all   $x\in  R\backslash  D$.

Let us tentatively assume that $f=g$ does not identically vanish   in $R^n \backslash D$.   The formal definition (1) (we use the principal value $p.v.$ abbreviation for the integral)  yields:
\be
(-\Delta)^{\alpha /2}f(x)= \mathcal{A}_{\alpha,n} p.v. \int_{R^n}
\frac{f(x)-f(y)}{|x-y|^{\alpha +n}}dy =  \mathcal{A}_{\alpha,n}
\left[ p.v. \int_{D}  \frac{f(x)-f(y)}{|x-y|^{\alpha +n}}dy  +   \int_{R^n\backslash D} {\frac{f(x)-g(y)}{|x-y|^{\alpha +n}}}dy \right] .
 \ee

Upon setting   $g(x)= 0$ for all   $x\in  R\backslash  D$,  we arrive at  the  restricted   fractional Laplacian $(-\Delta )^{\alpha /2}_D$,    whose integral definition is shared  with     $(-\Delta )^{\alpha /2}$, c.f. Eq. (1),   but   whose   domain   is restricted to  functions vanishing in $R^n\backslash D$.
Accordingly, for all $x\in D$  we have
\be
(-\Delta )^{\alpha /2}_D f(x) = (-\Delta )^{\alpha /2}f(x)= h(x)
\ee
where   the function $h(x)$   may not have  the exterior    property  of  $f(x)$. 	We  point out  that there is no restriction   upon the   the integration volume  which is a priori $R^n$  and not  solely  $D\subset R^n$.
Indeed:
\be
(-\Delta)^{\alpha /2}f(x)=  \mathcal{A}_{\alpha,n}
\left[ p.v. \int_{D}  \frac{f(x)-f(y)}{|x-y|^{\alpha +n}}dy  +   f(x)  \int_{R^n\backslash D} {\frac{dy}{|x-y|^{\alpha +n}}} \right] .
 \ee
so that the exterior  $R^n\backslash D$  contribution    affects the outcome of (7) for all $x\in D$.

In case of sufficiently regular domains, we encounter here a  solvable  spectral  (eigenvalue)  problem for the fractional Laplacian in  a  bounded domain $D$:
  $(-\Delta )^{\alpha /2}_D \phi _k(x)    = \lambda _k \phi _k(x)$, $k\geq 1$.
  This   spectral  issue has an ample coverage in the literature,  and with regard to a  detailed analysis of various eigenvalue problems  for   restricted fractional Laplacians, especially of the  fully computable  $1D$ case in  the   interval $(-1,1)=D \subset R$,    we mention  Refs.  \cite{kulczycki}, \cite{duo} - \cite{mypre}, see also \cite{kwa}.

 We note that Eq. (8)  can be converted to the form of the hypersingular Fredholm problem, discussed in detail in Refs.  \cite{zaba0,zaba1,mypre0,mypre}.
 All involved singularities can be properly handled (are removable) and  eventually one arrives at  the   formula,  \cite{mypre0}:
\be
(-\Delta )^{\alpha /2}_D f(x) \equiv  - {\cal{A}}_{\alpha,n}\int _{\bar{D}} {\frac{f(u)}{|u-x|^{n + \alpha }}}\, du
\ee
where $x\in D$ and $D$ is the ultimate  integration area. The
  $R\backslash D$  input,   implicit in  Eq. (8),  has been completely eliminated.

\subsubsection{Spectral fractional Laplacian}

 We first  impose the boundary conditions upon the Dirichlet  Laplacian in a bounded domain $D$  i. e. at the boundary $\partial  D$ of $D$.
 That is  encoded in the notation   $\Delta _{\cal{D}}$.  Presuming to have in hands its $L^2(D)$
  spectral solution (employed before in connection with (7)), we introduce a fractional power of the
  Dirichlet Laplacian $(- \Delta _{\cal{D}})^{\alpha /2}$ as follows:
  \be
  (-\Delta _{\cal{D}})^{\alpha /2}f(x) = \sum_{j=1}^{\infty } \lambda _j^{\alpha /2} f_j  \phi _j(x) =
  {\frac{1}{|\Gamma (-{\frac{\alpha }{2}})|}}\,
 \int_0^{\infty} (e^{t \Delta _{\cal{D}}}f - f) t^{-1-\alpha /2}\, dt .
  \ee
where  $f_j = \int _D f(x) \phi _j(x) dx$ and $\phi _j, j=1,2,...$ form an orthonormal basis
 system in $L^2(D)$:  $\int _D \phi _j(x) \phi _k(x) dx = \delta _{jk}$.

 We note that the spectral fractional Laplacian $(-\Delta _{\cal{D}})^{\alpha /2}$ and the  ordinary Dirichlet Laplacian $\Delta _{\cal{D}}$
 share eigenfunctions and their eigenvalues are  related as well: $\lambda _j \leftrightarrow \lambda _j^{\alpha /2}$.   The boundary data for
$(-\Delta _{\cal{D}})^{\alpha /2} $ are imported from these for  $\Delta _{\cal{D}}$.

 From the computational (computer-assisted)   point of view, this spectral simplicity has been considered as an advantage,
  compared to other proposals,  c.f. \cite{vazquez,teso,musina}.

 In contrast  to the situation in $R^n$,  the restricted  $(- \Delta )^{\alpha /2}_D$   and spectral  $(- \Delta _{\cal{D}} )^{\alpha /2}$
   fractional Laplacians  are inequivalent  and  have  entirely  different sets of   eigenvalues and eigenfunctions.
 Basic differences between them have been studied in \cite{servadei}, see also \cite{bucur,abtangelo}  and \cite{duo}. An extended study of the intimately related subordinate killed Brownian motion in a domain can be found in Refs. \cite{vondr,vondr1}.

We note one most obvious (and not at all subtle) difference encoded in the very definitions:  the boundary data for the restricted  fractional  Laplacian need to be exterior
and set on $R^n\backslash D$, while those for the spectral one are set locally  at the boundary $\partial D$  of $D$.

{\bf Remark 2:}  In the physics oriented research, the existence of inequivalent fractional Laplacians in case of bounded domains seems to have been  overlooked. That led to vigorious discussions  \cite{luchko,bayin}  about  the validity of spectral (eigenvalue problems) solutions presented in Refs. \cite{laskin0,laskin}.  See also \cite{wei}  for simple   standard quantum theory motivated problems, like e.g. the fractional   infinite well or harmonic oscillator with mass $m\geq 0$.   In particular, Laskin's spectral solution for the infinite well, as an outcome of his tacit assumptions, coincides precisely with the spectral Laplacian  eigenvalue solution $\lambda _k^{\alpha /2}, \phi _k(x), k\geq 1$, given the standard solution  $\lambda _k, \phi _k(x), k\geq 1$ of the quantum mechanical infinite well.
We point out  an explicit usage of the spectral fractional Laplacian in Ref. \cite{gitterman}, where the fractional process on the interval with absorbing ends has been resolved  by means of the spectral affinity with an infnite well problem,  \cite{gar,gar1,bickel}.

{\bf Remark 3:}
Let us mention that solutions of the  fractional infinite well problem,   together with that of an approximating sequence of deepening  fractional  finite wells,  based on the restricted fractional Laplacian,  have been addressed in Refs.  \cite{zaba,duo1,zaba0,zaba1,mypre}  and \cite{kwasnicki1,kaleta1,stos}.
Moreover, the fractional harmonic oscillator (including the so-called massless version. with the Cauchy generator involved)) has been addressed in a number of papers, \cite{gar0,zaba0,malecki} see also \cite{durugo} for the quartic case.

{\bf Remark 4:}  In Ref. \cite{buldyrev}  the spectral definition of the fractional Laplacian has been comparatively   mentioned, prior to the mathematically more refined  analysis of Ref. \cite{servadei}.  In fact, a quantitative  comparison has been made of the spectral and restricted Dirichlet  problems on the interval.  Numerical results for various average quantities  have been found  not to differ   substantially. It has been noticed that the restricted
Laplacian eigenfunctions  are close to the spectral Laplacian  eigenfunctions (trigonometric functions) except for the vicinity of the boundaries.  An important observation was that  the   probability  decay  time rate formulas show up detectable differences. In this connection it is also instructive  to record an analogous situation while  comparing the pure Brownian  case with its he spectral fractional  relative.

{\bf  Remark 5:} It is useful to exmplify the differences between the restricted Laplacian and spectral Laplacian outcomes for the interval $(-1,1)$. Namely, according to
\cite{kwasnicki1}, one arrives at an approximate eigenvalue formula for $n\geq 1$ and $0< \alpha  <2$: $
\lambda _n= \left[ {\frac{n\pi }{2}} - {\frac{(2-\alpha )\pi }{8}}\right]^{\alpha }  - O\left({\frac{2-\alpha }{n\sqrt{\alpha }}}\right)
$.
We note that the  eignevalues of the ordinary  (minus) Laplacian in the interval
read $\lambda _n= \left[ {\frac{n\pi }{2}}\right]^2$, $n\geq 1$.   Up to dimensional coefficients we have here the familiar quantum
mechanical spectrum of the infinite well,  set on the  interval  in question.
On the other hand, the spectral fractional well outcome is $\lambda ^{\alpha /2}_n= \left[ {\frac{n\pi }{2}}\right]^{\alpha }$, $n\geq 1$, see also \cite{laskin0,laskin,wei}.

{\bf Remark 6:}   The ground state function in the restricted case  has been proposed in  quite complicated  analytic  form (actrually it is an approximate expression  for a "true" eigenfunction),   \cite{kwasnicki1,stos}.  Porspective ground state  properties were also   analyzed  in minute detail,  with a  numemrical assistance, \cite{zaba0,zaba1,duo1}.  The available  data justify another approximation in terms of   a function  $
    \psi (x)= C_{\alpha ,\gamma } [(1-x^2) \cos(\gamma x)]^{\alpha /2}$,
where $C_{\alpha ,\gamma } $ stands for the $L^2(D)$ normalization factor, while $\gamma $ is considered to be the "best-fit" parameter, allowing to  approach quite closely the  computer-assisted eigenfunction outcomes, \cite{zaba0}.  This may be directly compared with the outcome for  the spectral case, whose  the ground state  $cos(\pi x/2)$   is shared with the ordinary  Laplacian in the interval.

\subsubsection{Regional fractional Laplacian}

The regional fractional Laplacian has been introduced in conjunction with the notion of censored symmetric  stable processes, \cite{bogdan}-\cite{barles}.
A censored stable process in an open set $D\subset  R^n$  is obtained from the
symmetric stable process by suppressing its jumps from  $D$  to the complement  $R^n \backslash D$
 of $D$, i.e., by restricting its L\'{e}vy measure to D.  Told otherwise,
 a censored stable process in an open   domain D is a stable process forced  to stay inside D. This makes a clear difference with a nuber of proposals to give meaning to Neumann-type  conditions, e.g. \cite{dipierro,abtangelo1}, where outside jumps are admitted, albeit with an immediate return ("resurrection", c.f. \cite{bogdan,pakes}) to  the interior of $D$.

Verbally the censorship idea resembles random  processes  conditioned to stay in a bounded domain forever, \cite{gar1,mazzolo}.
 However, we point out that the "censoring"  concept is not the same \cite{bogdan}  as that of the (Doob-type)   conditioning.
   Instead, it is intimately   related to  reflected stable processes   in a bounded domain with killing within the domain, or in the least  at its boundary,  encompassing  a class of  processes  (loosely interpreted as "reflective")  that do  not approach   the boundary at all, \cite{bogdan,refl}.

In Ref. \cite{refl} the reflected stable processes in a bounded domain have been investigated,  stringent criterions for their admissibility set  and their generators
identified with regional fractional Laplacians on the closed region $\bar{D}= D \cup \partial D$.
According to \cite{refl}, censored stable processes  of Ref. \cite{bogdan},  in $D$ and  for
$0<\alpha \leq 1$, are essentially the same as the reflected stable process.

In general, \cite{bogdan}, if $\alpha \geq 1$, the censored stable process  never approaches $\partial D$. If
$\alpha >1$, the censored process may have a finite lifetime and may take values at $\partial D$.

Conditions for the existence of the regional Laplacian for all $x\in \bar{D}$, in  spatial dimensions  higher than one,  have been set in     Theorem 5.3 of \cite{refl}.
For $1\leq \alpha <2$, the existence of the regional Laplacian  for all $x\in \partial D$, is granted if and only if  a  derivative   (this notion is not  conventional  and is adapted to the nonocal setting) of a each function in the domain in the inward normal direction vanishes, \cite{refl,warma}.

For our present purposes we assume $0<\alpha <2$ and $D\subset  R^n$ being an open set.
The regional Laplacian is assumed  (a technical assumption employed in the mathematical literature) to act upon functions $f$  on an open set $D$ such that
\be
\int _D {\frac{|f(x)|}{(1+ |x|)^{n+\alpha }}}  \,  dx < \infty
\ee
For such functions $f$, $x\in D$ and $\epsilon >0$, we  write  (compare e.g. Eqs. (6) - (9))
\be
(-\Delta )^{\alpha /2}_{D,Reg}  f(x)  = \mathcal{A}_{\alpha,n}
\lim\limits_{\varepsilon\to 0^+}
\int\limits_{y\in D \{|y-x|>\varepsilon\}}
\frac{f(x)-f(y)}{|x-y|^{\alpha +n}}dy.
\ee
provided the limit (actually the Cauchy  principal value, p.v.) exists.

Note a subtle difference between the restricted  and regional fractional Laplacians.
The former is restricted exclusively  by the domain property $f(x)=0, x\in R^n\backslash D$. The latter
is   restricted exclusively   by demanding the  integration variable $y$  of the L\'{e}vy measure  to be in $D$.

If we impose  the   Dirichlet  domain restriction upon the regional  fractional  operator   ($f(x)=0$ for $x\in R^n\backslash D$, $D$ being  an open set $D$),   we can rewrite Eq. (8)   as an  identity  relating the restricted and regional fractional Laplacians  for all $x\in D$, \cite{bogdan,duo,gar}:
 \be
 (- \Delta )^{\alpha /2}f(x) =  \left[(-\Delta )^{\alpha /2}_{D,Reg}    + \kappa _D(x)\right] f(x)
\ee
where $
\kappa _D(x) =    \mathcal{A}_{\alpha,n}  \int_{R^n\backslash D} {|x-y|^{-(n+\alpha )}} dy$
is nonnegative  and  plays the role of the density of the killing measure,     \cite{bogdan}.

 Eqs. (12), (13), actually indicate that  the restricted fractional Laplacian   can be  obtained as a $\kappa _D$ perturbation of the regional one (see e.g. \cite{bogdan}) and in principle we can move a priority status from the restricted to the regional one. Provided the exterior  Dirichlet condition is respected by both operators.

Indeed,  while quantifying the random process   associated with  the generator  $(-\Delta )^{\alpha /2}_{D,Reg} +  \kappa _D(x) $ by invoking the concept of the    Feynman-Kac kernel \cite{bogdan},   one is tempted to  view   $\kappa _D(x)dt$ as a probability of extiction (killing) in the time interval $[t,t+\Delta t]$, and accordingly $\kappa _D(x)$ stands for the killing rate.

Proceeding otherwise, we can define the regional Laplacian as a perturbation of the restricted fractional one by the negative (not positive-definite)  potential:
$(- \Delta )^{\alpha /2} - \kappa _D(x)=(-\Delta )^{\alpha /2}_{D,Reg}$.   Then however,  $\kappa _D(x)$ would refer to the resurrection (birth,  or  creation, \cite{bogdan}) rate.
One should be aware, that to give meaning to the regional fractional Laplacian,  various precautions concerning its domain must be observed to handle potentially dangerous divergent terms  ($\kappa _D(x) $ being included).

If  to replace $D$   in Eq. (12)  by  $\bar{D}=D\cup \partial D$ then,  according to \cite{bogdan,refl},    one arrives at the  generator of  a reflected stable process  in $\bar{D}$, c.f. \cite{refl},  $(-\Delta )^{\alpha /2}_{\bar{D},Reg} f(x) $. Provided suitable  conditions (various forms of  the  H\"{o}lder continuity) upon functions in the domain of the nonlocal operator are  respected.
In particular, in case of $1 \leq \alpha < 2$ it has been shown that $(-\Delta )^{\alpha /2}_{\bar{D},Reg} f(x) $
exists at  a boundary   point $x \in  ∂D$  if and only if   the normal inward derivative vanishes: $(\partial f/\partial n)(x) = 0$.

We note that an existence of the spectral solution (eigenvalue problem) for the regional Laplacian with (appropriately defined, generally not in an appealing  derivative form) Neumann boundary data has been demonstrated in Ref. \cite{warma}, see e.g. also for the one-dimensional  discussion in \cite{bogdan, guan}.\\

{\bf Remark 7:}  To avoid possible misunderstandings and misuse of the  concept of killing (and subsequently that of resurrection/birth,   \cite{pakes,evans,evans1,durang}), let us recall  basic Brownian  motion intuitions that underly the the implicit  path integral formalism.  Namely,  operators of the form $\hat{H}= - (1/2)\Delta + V  \geq 0$  with $V\geq 0$  give rise to transition kernels of diffusion -type  Markovian processes with killing (absorption), whose rate is determined
 by the value of $V(x)$ at $x\in R$. That interpretation stems from the celebrated Feynman-Kac (path integration) formula, which assigns to $\exp(-\hat{H}t)$
 the positive  integral kernel
\[
  \left [\exp(- (t-s)(-{\frac{1}2}\Delta  + V)\right](y,x)  =
       \int   \exp\left[-\int_s^t V(\omega(\tau  )) d\tau \right]\,   d\mu _{s,y,x,t}(\omega)
 \]
 In terms of Wiener paths that kernel is constructed as a path integral over paths that get killed at a point $X_t=x$,   with an extinction  probability $V(x)dt$     in the time interval$(t,t+dt)$.     The killed path is henceforth removed from the ensemble of on-going Wiener paths.   The exponential factor $\exp[-\int_s^t V(\omega(\tau  ))$  is here responsible for a proper redistribution of Wiener paths, so that  the evolution rule
\[
\big([\exp (t L)]f\big)(x)  = \int _{R^n} k(x,0;y,t) f(y) dy   =  E^x \big[f(X_t)\big]
\]
 with $-L= -(1/2)\Delta  + V$ is well defined as an expectation value of the killed process  $X(t)$,  but given in terms of Brownian paths $W(t)$ with the   Feynman-Kac   weight:
 $$E^x [f(X_t)] = E^x\left[f(W_t) \exp \left(-\int_0^t  V(W_\tau )d\tau \right)\right].$$
 The resurrection  executed with the same $V(x)$   rate would actually refer to another Feynman-Kac weight (note the sign change): $\int   \exp[\int_s^t V(\omega(\tau  )) d\tau ]\,   d\mu _{s,y,x,t}(\omega)$, which amounts to  the replacement of $V$ by $-V$  in the expression for the motion   generator   $L$  (as  an instructive example one may conceive a replacement of the harmonic oscillator potential by its inverted version, c.f. \cite{dual}).

\section{Relevance versus irrelevance of eigenvalue problems for fractional Laplacians in  a bounded domain:  Stochastic viewpoint.}

\subsection{Transition densities}

Let us restate  our motivations  in a more formal lore  (our notation is consistent with that in Ref. \cite{lorinczi}).
  Namely, given the  (negative-definite) motion  generator $L$,  we shall
 consider  the (contractive)  semigroup  evolutions   of the form
\be
f(x,t) = T_tf(x)=   \big([\exp (t L)]f\big)(x)  = \int _{R^n} k(x,0;y,t) f(y) dy   =  E^x [f(X_t)], \ee
 where  $t\geq 0$.
In passing, we have  here  defined a local expectation value  $E^x[...]$, interpreted as  an average   taken at time  $t>0$,
  with  respect  to the process $X_t$ started in $x$ at $t=0$,  with values  $X_t=y \in R^n$  that are  distributed
  according to the  positive  transition  (probability)  density  function  $k(x,0;,y,t)$.

  We in fact deal with a bit more  general transition function
     $k(x,s;y,t), 0\leq s<t$ that is symmetric with respect to $x$ and $y$, and  time homogeneous. This  justifies
  the  notation $k(x,s;y,t)= k(t-s,x,y)=k(t-s,y,x)$ and subsequently  $k(x,0;y,t)=  k(t,x,y)=k(t,y,x)$.   The "heat" equation  $
\partial _t f(x,t)= L f(x,t)$  for $t \geq 0$  is    here   presumed to  follow.  We recall, that given a  suitable  transition function, we
 recover the semigroup   generator via   $
  [Lf](x)= \lim_{t\rightarrow 0} {\frac{1}{t}}  \int_{R^n}
 [k(t,x,y) f(y)dy - f(x)]$,
in accordance with an (implicit strong continuity) assumption that  actually  $T_t=\exp(Lt)$.

For completeness let us mention that the semigroup property $T_t  T_s =T_{t+s}$,  implies
the validity of  the composition rule    $\int_{R^n} k((t,x,y)\, k(s,y,z)\, dy = k(t+s,x,z)$.
 Let $B\subset R^n$, a probability that a subset $B$ has been reached by the process  $X_t$   started  in  $x\in R^n$,   after the
 time lapse  $t$,  can be inferred from   $ P[X_t \in B|X_s=x]= \int_B k(t-s,x,y)dy$, $0\leq s<t$, and reads
\be
 P^x(X_t\in B) = \int_B k(t,x,y)dy = k(t,x,B)
\ee
Clearly, $P^x(X_t\in R^n)=1$.

In general,  for time-homogeneous processes,   we have  $k(x,s;B,t) = \int _B k(t-s,x,y) dy$,   $s<t$,
hence  we can rephrase the Chapman-Kolmogorov relation as  follows:
 \be
 \int_{R^n} k(x,s;z,u) k(z,u,B,t) dz = k(x,s;B,t)= k(t-s,x,B)=  P[X_{t-s} \in B|X_s=x],
 \ee
 where $s<u<t$.

\subsection{Absorbing boundaries and survival probability}

Now,  we shall pass  to killed Brownian and L\'{e}vy-stable motions in a bounded domain.
Let us denote $D$ a bounded open set in $R^n$. By $T_t^D$ we denote  the semigroup  given by the process $X_t$ that is killed on exiting $D$.
Let  $k_D(t,x,y)$ be the transition density for $T_t^D$. Then  \cite{kulczycki}:
\be
T_t^Df(x)= E^x[f(X_t); t <\tau _D] =  \int_D  k_D(t,x,y) f(y) dy
\ee
provided  $x\in D$ , $t>0$ and   the first exit time $\tau _D= \inf \{t \geq 0, X_t \notin D \}$  actually stands for the  killing time  for $X_t$.

From the general theory of killed semigroups in a bounded domain there follows that in  $L^2(D)$ there exists  an  orthonormal    basis of eigenfunctions
$\{ \phi _n \}, n=1,2,...$ of $T_t^D$ and corresponding eigenvalues $\{ \lambda _n, n=1,2,... \}$  satisfying   $0<\lambda _1 <\lambda _2 \leq \lambda _3 \leq ...$.
Accordingly there holds $T_t^D \phi _n(x) = e^{-\lambda_nt}\, \phi _n(x)$, where  $x\in D, t>0$ and we also have:
\be
k_D(t,x,y) = \sum_{n=1}^{\infty } e^{-\lambda_nt}\, \phi _n(x)\, \phi _n(y)
\ee
The eigenvalue $\lambda _1$ is non-degenerate (e.g. simple)  and the corresponding strictly positive   eigenfunction  $\phi _1$ is often called
  the ground state function.

   For the infinitesimal generator   $L_D$  of the semigroup we have  $L_D \phi _n(x)= -\lambda _n \phi _n(x)$
  The corresponding "heat" equation $\partial _t f(x,t)= L_D f(x,t)$  holds true  as well.

   It is useful to introduce the notion of the survival probability for the killed random process in a bounded domain  $D$.
   Namely, given $T>0$, the probability that the random motion has not yet been absorbed (killed) and thus survives up to time $T$  is given by
   \be
   P^x[\tau >T] = P^x[X_T \in  D] = \int_D k_D(T,x,y) dy
\ee
and is  named  the survival probability up to time $T$.

Proceeding  formally  with  Eqs. (4) and (5), under suitable  integrability  and   convergence assumptions for the infinite series, one arrives at an asymptotic survival  probability  decay rule:
\be
P^x[\tau >T]= \sum_{n=1}^{\infty } e^{-\lambda_nT}\, a_n \,  \phi _n(x) \,  \Rightarrow \,   a_1\,  e^{-\lambda_1T}\, \phi _1(x)
\ee
where $a_n = [\int_D  \phi _n(y) dy] $, $n=1,2,...$.   This  familiar exponential decay law  is   characteristic for   e.g. the
 Brownian  motion with    absorbing boundary data.  Its time rate is  controlled by  the largest  eigenvalue  $ -\lambda _1$ of $\Delta _D$.

 Thus, as far as the asymnptotic properties are concerned, we do not need the fully-fledged spectral solution (e.g. that of the eigenvalue problem for the fractional Laplacian). The lowest eigenvalue and its eigenfunction are what really matters.
 The  degree of accuracy with which we approximate  the asymptotic behavior may be improved by not ignoring the second eigenvalue (i.e. the fundamental gap, \cite{gaps}). That is  also specific  to the reflected motion with the lowest eigenvalue zero.

\subsection{Conditioned random motions in a bounded domain: taboo processes.}

For  the absorbing stochastic process with the transition density (17) (thus  surviving up  to time $T$), we  introduce
  survival probabilities  $P^y[\tau > T-t]$  and $P^x[\tau >T]$,  respectively  at times $T-t$ and $T$,  $0<t<T$.
 We infer a   conditioned  stochastic  process  with the transition density:
 \be
 q_D(t,x,y) = k_D(t,x,y) {\frac{P^y[\tau > T-t]}{P^x[\tau >T]}},
 \ee
 which   by construction   survives up to time  $T$  and  is additionally conditioned to start in  $x\in D$ at time $t=0$  and reach the
  target   point  $y\in D$, at time $t<T$.   An  alternative   construction  of such processes, in the diffusive case,  has been described  in \cite{gar1}.

 Given $t<T$, in the large time asymptotic of T, we can invoke (19), and  once  $T \rightarrow  \infty $ limit is executed, Eq.  (20) takes the form:
 \be
q_D(t,x,y) \longrightarrow   p_D(t,x,y) = k_D(t,x,y) {\frac{ \phi _1(y)}{ \phi _1(x)}}  \exp(\lambda _1 t)
 \ee
We have arrived at the  transition probability density  $p_D(t,xy)$  of the   probability conserving  process,  which never leaves the  bounded  domain $D$.
The latter "eternal life-time" property is shared with the  censored processes of  Ref. \cite{bogdan}, but the pertinent taboo processes  (c.f.  \cite{gar1,mazzolo,pinsky}
 for  the origin of the term "taboo") appear to form an independent family of random motions.

 Its   asymptotic (invariant) probability density   is $\rho (y) =[\phi _1(y)]^2$,  $\int_D \rho (y)\, dy =1$
  (that in view of  the implicit $L^2(D)$ normalization  of eigenfunctions $\phi _n$).

 By  employing (19)  and the definition  $\rho (y) =[\phi _1(y)]^2$,   we  readily check the stationarity property. We take  $\rho (x)$ as the initial
 distribution (probability density) of   points in  which the process is started  at time $t=0$).
 The  propagation towards target points,  to be reached at  time $t>0$,   induces a  distribution $\rho (y,t)$. Stationarity follows from:
\be
\rho (y,t)=  \int_D  \rho (x) p_D(t,x,y) dx = \rho (y).
\ee
Note that  in contrast to  $k_D(t,x,y)$ the transition probability function $p_D(t,x,y)$ is no longer a symmetric function of $x$ and $y$.  Compare e.g. in this connection Ref. \cite{kaleta}.

  {\bf Remark 8:}
    The semigroup $T_t^{D}(\alpha ) = \exp(- t(-\Delta )^{\alpha /2}_D$, $t\geq 0$  of the stable process killed
upon exiting from a  bounded   set $D$ has an eigenfunction expansion of the form (19). Basically we never have in
hands a complete set of eigenvalues and eigenfunctions and likewise we generically do not know a closed analytic
form for the semigroup kernel $k_D(t,x,y)$, (19). A genuine mathematical achievement has been to establish
that when  $\alpha \in (0,2)$  and  a bounded domain $D$ is a subset of $R^n$, then  the stable semigroup
  $T_t^{D}(\alpha )$  is intrinsically ultracontractive. This technical (IU) property actually means that
  for any $t>0$ there exists  $c_t$ such that  for any $x,y \in D$  we have, \cite{kulczycki,kulczycki1,grzywny,chen}:
  $$
k_D(t,x,y) \leq c_t\, \phi _1(x) \phi _1(y).
  $$
Actually we have $k_D(t,x,y) = \sum_1^{\infty } e^{-\lambda _nt} \phi _n(x) \phi _n(y)$.  Accordingly:
$$
 {\frac{k_D(t,x,y)}{e^{-\lambda _1 t} \phi _1(x) \phi _1(y)}}= 1 +
 \sum_2^{\infty } e^{-(\lambda _n- \lambda _1)t}
  \, {\frac{\phi _n(x) \phi _n(y)}{\phi _1(x) \phi _1(y)}}.
$$

It follows that we have a complete information about the (large time asymptotic) decay of
relevant quantities:
$$
\lim_{t\to \infty }{\frac{k_D(t,x,y)}{e^{-\lambda _1 t} \phi _1(x) \phi _1(y)}} =1
$$
and  (for $t>1$)
$$
e^{-(\lambda _2- \lambda _1)t}  \leq  \sup_{x,y \in D}  | {\frac{k_D(t,,x,y)}{e^{-\lambda _1 t}
\phi _1(x) \phi _1(y)}}|
\leq C_{\alpha ,D} e^{-(\lambda _2 - \lambda _1)t}.
$$
Thus, what we actually need to investigate the large time regime of L\'{e}vy processes in the
bounded domain $D$, is to know two lowest eigenvalues $\lambda _1, \lambda _2$  and the ground state eigenfunction  $\phi _1(x)$ (eventually its shape, \cite{banuelos}) of the motion generator, see e.g. also \cite{gaps}.
 The existence of conditioned L\'{e}vy  flights, with a transition density (21) and  an invariant probability density $\rho (y) =[\phi _1(y)]^2$,  $\int_D \rho (y)\, dy =1$ is here granted.

\subsection{Reflected motions in a bounded domain}

Reflected random motions in the bounded domain are typically expected to live indefinitely,  never leaving the domain,   basically with a complete reflection
form the boundary.  (We cannot a priori exclude  a partial reflection, that is accompanied by  killing or transmission.)

In case of previously considered motions a boundary  may be regarded as  either
a transfer terminal to the so-called "cemetry' (killing/absorption), or  as being  inaccessible  form the interior  at all  (conditioned processes).
In both scenarios, the major technical tool was the eigenfunction expansion (11), where the spectral solution for the Laplacian with the
 Dirichlet boundary data has been employed.  Thus,   in principle   we should here use the notation $\Delta _{\cal{D}}$, where
 ${\cal{D}}$ indicates  that the  Dirichlet boundary data   have been  imposed at the boundary  $\partial D$  of  $D\subset R^n$.

 Reflecting boundaries are related to Neumann boundary data, and therefore  we should rather use the notation $\Delta _{\cal{N}}$.
 In a bounded domain we deal with  a spectral (eigenvalue) problem for $\Delta _{\cal{N}}$ with  the  Neumann data-respecting  eigenfunctions
 and eigenvalues.

  The major difference, if compared to the absorbing case is that the eigenvalue zero is  admissible and the corresponding
 eigenfunction  $\psi_0(x)$ determines an asymptotic (stationary, uniform in $D$) distribution  $\rho _0(x)= [\psi _0(x)]^2$, \cite{bickel0,bickel}.
In the Brownian context, the rough  form of the related  transition density   looks like:
\be
k_{\cal{N}} (t,x,y) = {\frac{1}{{\mathrm{vol}}(D)}}  +  \sum_{n=1}^{\infty } e^{-\kappa_nt}\, \psi _n(x)\, \psi _n(y)
\ee
where   $\kappa _n$   are positive eigenvalues,  $\psi _n(x)$ respect the Neumann boundary data and $vol(D)    $  denotes
 the volume  of $D$ (interval length, surface are etc.).  We have $\psi_0(x)= 1/\sqrt{{\mathrm{vol}}(D)}$.

 If one resorts to the spectral definition of the fractional Laplacian in a bounded domain,  Neumann conditions are directly imported from these for the standard Laplacian, and thence (see e.g. \cite{gitterman,buldyrev, bickel}  the Laplacian  eigenvalues     $\lambda _k, k>0$   need to be replaced by $\lambda _k^{\alpha /2}$. That is enough to map an eigenfunction expansion of the  Laplacian-induced pdf  (probability density function) and transition pdf to these appropriate for the fractional case (remebering about the spectral definition of the fractional Laplacian, see also \cite{servadei}).

 {\bf Remark 9:} Let us invoke the standard Laplacian in the interval (with
 an immediate  passage to the  spectral definition in  mind).  The case
 of reflecting boundaries  in the interval   is specified by Neumann boundary conditions     for solutions of the diffusion  equation
$\partial _tf (x) = \Delta _{\cal{N}}f(x)$  in the interval $\bar{D}= [a,b]$.  We  need to have respected   $(\partial _x f) (a)= 0 =   (\partial _x f)(b)$ at
the interval boundaries.
  The pertinent transition density   reads, \cite{carlsaw,bickel}:
  \be
k_{\cal{N}} (t,x,y)  =     {\frac{1}{L}}  +  {\frac{2}L} \sum_{n=1}^{\infty } \cos\left( {\frac{n\pi
}L}(x - a)\right) \cos\left( {\frac{n \pi }L} (y - a)\right)  \exp
\left(- {\frac{n^2\pi ^2}{L^2}} \, t  \right).
\ee

The operator $\Delta _{\cal{N}}$ admits the eigenvalue $0$ at the bottom of its spectrum, the corresponding eigenfunction being a constant whose square  actually  stands for  a
 uniform probability distribution on the interval of length $L$ ($L=2$ in case of  $(-1,1)$),   to be approached in  the  asymptotic  (large time)  limit.
 Solutions of the diffusion equation with reflection at the boundaries of $D$  can be modeled  by setting $p(x,t)= k_{\cal{N}}(t,x,x_0)$, while remembering that $p(x,0)=
 \delta(x-x_0)$.  We can as well resort to  $c(x,t)= \int_D k_{\cal{N}} (t,x,y) c(y) dy$, while keeping in memory that $k(t,x,y)= k(t,y,x)$.
Note that all $n\geq 1$ eigenvalues coincide with these of the absorbing case, \cite{gitterman}.  The  eigenvalues of the spectral fractional operator for $n>0$
 would read $\lambda _n^{\alpha /2}=
(n^2\pi ^2  /4)^ {\alpha /2}= (n\pi /2)^{\alpha }$.  For comparison we reproduce  a rough  analytic  approximation of the restricted Laplacian eigenvalues $\left[ {\frac{n\pi }{2}} - {\frac{(2-\alpha )\pi }{8}}\right]^{\alpha }$,   \cite{kwasnicki1}, see also \cite{mypre} for an extended list of  numerically   generated sharp  eigenvalues.

\section{Spectral properties  of the regional  fractional
 Laplacian.}

 All derivations and mathematically advanced demonstrations of the existence of regional fractional Laplacianas and their link with  (whatever that actually means) reflected        L\'{e}vy flights and general censored processes, involve subtleties concerning the behavior of functions in their domain,  in the vicinity and at the boundaries.  Attempts to reconcile a semi-phenomenologically  prescribed  boundary behavior of   censored stochastic processes with the notion of  (i) regional fractional Laplacian and  (ii) any conceivable forms of Neumann boundary data,  lead to inequivalent outcomes.  That refers as well   to the exoploitation of Sobolev spaces  instead of the  more familiar for physicists $L^2(D)$ Hilbert   spaces, \cite{warma,guide}.
 Other subtleties,  like an issue of the H\"{o}lder continuity up to the boundary,  need to be   kept under control as well.

All that blurs or even  hampers   any  pragmatic  approach aiming
   at  the deduction of    approximate  (if not sharp)  eigenvalues,    functional shapes   of eigenfunctions,   related    (approximate)  probability distributions and estimates of
probability killing rates or decay rates towards  equilibrium, if  regional fractional Laplacians  are interpreted as primary theoretical constructs, instead of the restricted ones.

It is our purpose to  follow the  pragmatic routine, tested before in our investigation of fractional Laplacians with exterior  Dirichlet boundary  conditions, \cite{zaba0,zaba1,mypre}.  We point out the existence of alternative  computation routines (based on a direct  discretization of  fractional Laplacians), \cite{duo1,duo2,pezzo}.  It is advantageous that we can directly compare our results with these of Ref. \cite{duo}  for the regional Laplacian with  the   (exterior)   Dirichlet boundary data.
THis test of a comparative validity of two different computation methods gived support to our subsequent analysis of the full-fledged spectral problem with no Dirichlet restriction, resulting in the  implicit relecting boundary  conditions.

It is worthwhile to note that  the main body of the mathematical research refers to space dimensionalities $d\geq 2$.   One should keep in mind, in the least for  comparative purposes, that the  one-dimensional case  (being of interest for us in below)   differs form higher dimensional ones  in a number of technical points. Notwithstanding the one dimensional case has an advantage of computability and allows for a deeper insight into the   properties of  motion generators and processes, specifically into an issue of potentially dangerous divergencies.

\subsection{Restricted versus regional  fractional Laplacian}

We depart from the formal definition (11)-(13) of the regional fractional Laplacian and following \cite{gar,duo}  concentrate on the one-dimensional spectral problem  in the interval.  The  latter  topic, for   the Dirichlet fractional Laplacians,  has been  addressed before in a number of papers   in the whole stability parameter range $0<\alpha < 2$ and in more  detail,  for the distinguished  $\alpha =1$  value (Cauchy process  and motion generator) in Refs. \cite{dybiec,dybiec1,dubkov,denisov},  \cite{zaba,zaba0,zaba1,duo1,mypre}  and \cite{kwasnicki,malecki}.

We consider the  regional fractional Laplacian (see, e.g. Ref. \cite{refl}) acting on the  closed interval (including the endpoints)  $-1 \leq x \leq 1$. To make  clear   the difference between the regional and ordinary fractional Laplacians,   we begin with  the  fractional Laplacian on the whole real axis. We have, \cite{mypre}:
\begin{equation} \label{ger1}
(-\Delta )^{\alpha /2}\psi(x)=-A_\alpha  \int_{-\infty}^\infty \frac{\psi(u)-\psi(x)}{|u-x|^{1+\alpha }}du,\ A_\alpha =\frac{1}{\pi}\Gamma(1+\alpha )\sin\frac{\pi \alpha }{2}.
\end{equation}

Since we assume the exterior Dirichlet condition  $\psi(u)=0$   for $R\backslash (-1,1)$  (c.f. (8), (9)),  the division of the integral into separate contributions from within and ouside the interval
\begin{equation} \label{ger2}
(-\Delta )^{\alpha /2}\psi(x)=-A_\alpha  \int_{-\infty}^\infty \frac{\psi(u)-\psi(x)}{|u-x|^{1+\alpha }}du=-A_\alpha \left[\int_{-\infty}^{-1}+\int_{-1}^1+\int_1^\infty\right]\frac{\psi(u)-\psi(x)}{|u-x|^{1+\alpha }}du.
\end{equation}
implies:
\begin{eqnarray}
&&(- \Delta )^{\alpha /2}\psi(x)=-A_\alpha \left[-\psi(x)\int_{-\infty}^{-1}\frac{du}{|u-x|^{1+\alpha }}+\int_{-1}^1\frac{\psi(u)}{|u-x|^{1+\alpha }}du-\psi(x)\int_{-1}^1\frac{du}{|u-x|^{1+\alpha }}-\psi(x)\int_1^{\infty}\frac{du}{|u-x|^{1+\alpha }}\right]\equiv \nonumber \\
&&\equiv -A_\alpha \left[\int_{-1}^1\frac{\psi(u)}{|u-x|^{1+\alpha }}du-\psi(x)\int_{-\infty}^\infty \frac{du}{|u-x|^{1+\alpha }}\right]\equiv -A_\alpha  \int_{-1}^{1}\frac{\psi(u)du}{|u-x|^{1+\alpha }}+\Delta I_{1D}. \label{ger3}
\end{eqnarray}
Formally, the integral $\Delta I_{1D}$ in \eqref{ger3}  indentically vanishes
\begin{equation} \label{ger4}
\Delta I_{1D}=-\frac{A_\alpha  \psi(x)}{\alpha }\left[\frac{1}{|u-x|^\alpha }\right]_{-\infty}^\infty=0
\end{equation}
for all  $0<\alpha <2$. This implies that the eigenvalue problem for the operator $|\Delta|^{\alpha /2}$ in the interval  (i.e. actually   the restricted fractional Laplacian  $(-\Delta )^{\alpha /2}_D$ of Eq. (9))   is defined by the equation
\begin{equation} \label{ger5}
(-\Delta )^{\alpha /2}\psi(x) =  -A_\alpha  \int_{-1}^{1}\frac{\psi(u)du}{|u-x|^{1+\alpha }}=E\psi(x).
\end{equation}
 The emergent spectrum   is discrete  $E\equiv E_n$, $n=1,2...$, non-degenerate  and positive $E_n>0$, see e.g. \cite{kulczycki,mypre0,mypre} and  references therein.

The regional fractional Laplacian is defined similarly to Eq. \eqref{ger1} but  directly  in the interval  $D= [-1,1]$ (that refers to the admitted integration area as well), i.e.
\begin{equation} \label{er1}
(-\Delta )^{\alpha /2}_{D, reg} \psi(x)=-A_\alpha  \int_{-1}^1 \frac{\psi(u)-\psi(x)}{|u-x|^{1+\alpha }}du=-A_\alpha \left[\int_{-1}^{1}\frac{\psi(u)du}
{|u-x|^{1+\alpha }}-\psi(x)\int_{-1}^{1}\frac{du}{|u-x|^{1+\alpha }}\right].
\end{equation}

We can readily evaluate the integral:
\begin{equation}\label{er2}
\int_{-1}^{1}\frac{du}{|u-x|^{1+\alpha }}=-\frac{1}{\alpha }\frac{1}{(1-x^2)^\alpha }\bigg[(1-x)^\alpha +(1+x)^\alpha \bigg]  = - {\frac{1}{A_{\alpha }}} \kappa_{D}^{\alpha }(x)
\end{equation}
where $D=[-1,1]$ and  we meet  essentialy  the same  $\kappa _D(x)$ that has emerged in the formula (13), see also Eq. 2.13 in Ref. \cite{duo}. We emphasize that apart from  a difference in the integration volumes for the involved integrals  (13) and (32), the outcome of integrations is the same, see  e.g.  also \cite{mypre}).
We note a familiar   \cite{mypre} outcome  $2/(x^2-1)$   for  $\alpha =1$.

\begin{figure}[tbh]
\centering
\includegraphics[width=0.7\columnwidth]{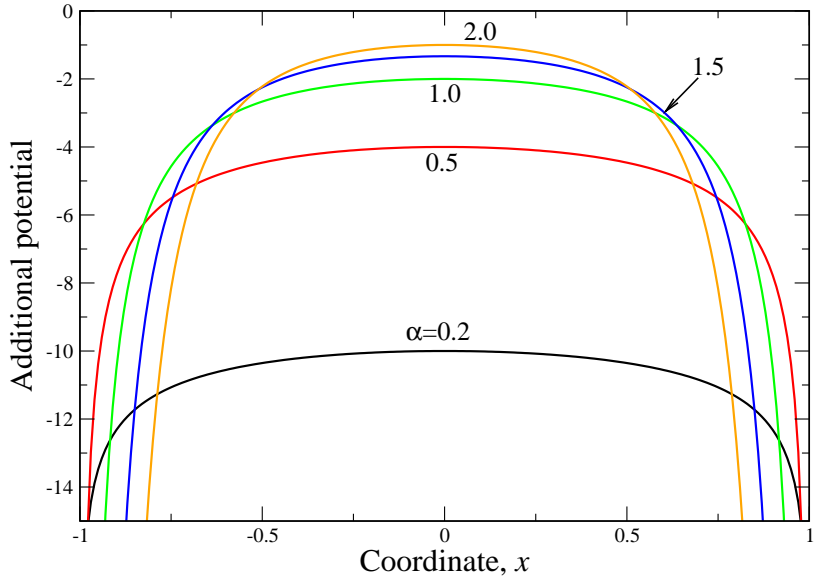}
\caption{The negative-definite singular  potential \eqref{er2} in the spectral problem for the  regional fractional Laplacian for  various values of  $\alpha $.} \label{ig1}
\end{figure}

We  thus arrive at  the  following spectral problem for the regional fractional Laplacian  in $D=[-1,1]$:
\begin{equation}\label{er3}
-A_\alpha  \int_{-1}^{1}\frac{\psi(u)du}{|u-x|^{1+\alpha }}-\frac{A_\alpha  \psi(x)}{\alpha }\frac{(1-x)^\alpha +(1+x)^\alpha }{(1-x^2)^\alpha }=E\psi(x).
\end{equation}
In short that reads
\be
 (-\Delta )^{\alpha /2}_{D,Reg}\psi (x) =  (-\Delta )^{\alpha /2}\psi(x)  - \kappa_{D}^{\alpha }(x)  \psi (x)= E\psi(x)
 \ee
with the persuasive interpretation of  $(-\Delta )^{\alpha /2}_{D,Reg}\psi (x)$  as   the additive perturbation of $ |\Delta|^{\alpha /2}$ by the   negative-definite  potential  $ - \kappa_{D}^{\alpha }(x)$. The  latter is  an inverted version  (c.f. \cite{dual})  of the attractive singular  (c.f. \cite{diaz}) potential  $\kappa _{D}^{\alpha }(x)$.   Its   functional   shape (the coefficient $A_{\alpha }$ has been  skipped)  for different  values of  $\alpha $ is reported in Fig.\ref{ig1}.

In passing we note  that a concept of resurrected  (after  killing) Markov processes  has been associated with so interpreted random noise generator  $(-\Delta )^{\alpha /2}_{D,Reg}\psi (x)$,  see e.g. \cite{bogdan} and \cite{pakes}.

\subsection{Regional fractional Laplacian  in $L^2(D), D=[-1,1]$. Trigonometric base with Dirichlet boundary conditions.}

We do not know of  any   methods  towards an analytic solution of  the pertinent eigenvalue problem  and  therefore  we  reiterate to numerically-assisted arguments, where an explicit diagonalization of  Eq.  \eqref{er3}   in the $L^2(D)$   trigonometric basis can  be performed.
To this end we  shall  use the same "even-odd" base (and the computation method)  as that in Ref. \cite{mypre}.   For any function  $\psi _{\alpha}(x)$,    that  may  possibly be  a solution to the eigenvalue problem,  we have the folowing   expansions in the trigonometric basis of $L^2(D), D=[-1,1]$:
\begin{equation} \label{eod}
\psi_{\alpha  e}(x)=\sum_{k=0}^\infty a_{k\alpha }\cos \frac{(2k+1)\pi x}{2},\ \psi_{\alpha  o}(x)=\sum_{k=1}^\infty b_{k \alpha } \sin k\pi x.
\end{equation}
So defined would-be  basis functions \eqref{eod}  satisfy the  Dirichlet boundary conditions $\psi_{\alpha  e,o}(\pm 1)=0$.  The behavior of the derivative $d\psi(x=\pm1)/dx$  is immaterial.

Passing to the matrix representation of the eigenvalue problem, for  the  even subspace we have
\begin{equation}\label{ev1}
\sum_{k=0}^\infty a_k (\gamma_{ki\alpha }+\beta_{ki\alpha })=Ea_{i\alpha }.
\end{equation}
Explicitly, the matrix to be diagonalized has the form (we suppress an  index $\alpha $ for a moment)
\begin{equation}\label{so1}
\hat A_{even}=\left(\begin{array}{cccc}\gamma_{00}+\beta_{00}  & \gamma_{10}+\beta_{10}  & \cdots & \gamma_{n0}+\beta_{n0} \\
\gamma_{10}+\beta_{10} &\gamma_{11}+\beta_{11} & \cdots & \gamma_{n1}+\beta_{n1}  \\
 \vdots & \cdots & \cdots & \vdots \\
\gamma_{n0}+\beta_{n0} & \gamma_{n1}+\beta_{n1}  & \cdots & \gamma_{nn}+\beta_{nn}
\end{array}\right),
\end{equation}
where $\gamma_{ik}\equiv \gamma_{ki}$ are "old" \cite{mypre} matrix elements coming from Eq. \eqref{ger5} (or first term in left hand side of Eq. \eqref{er3}):
\begin{equation}\label{so2}
\gamma_{ik\alpha }=\int_{-1}^1 f_{k\alpha }(x)\cos \frac{(2i+1)\pi x}{2}dx,
\end{equation}
where $f_{k\alpha }(x)$ are defined by Eq. (16) of Ref.  \cite{mypre}.

At the same time, $\beta_{ki}$ are "new" matrix elements coming from the second term in the left-hand side of \eqref{er3}, i.e. from the additive negative-definite
 potential \eqref{er2}:
\begin{equation} \label{so3}
\beta_{ik\alpha }=-\frac{A_\alpha }{\alpha } \int_{-1}^1 \cos \frac{(2i+1)\pi x}{2} \cos \frac{(2k+1)\pi x}{2}\ \frac{(1-x)^\alpha +(1+x)^\alpha }{(1-x^2)^\alpha }dx.
\end{equation}
The integrals for $\beta_{ik1}$ ($\alpha =1$) can be calculated explicitly
\begin{equation}\label{so4}
\beta_{ik1}=\frac{1}{\pi}\Bigg\{(-1)^{k-i}{\mathrm {Ci}}[2\pi(1+k+i)]+(-1)^{k+i+1}{\mathrm {Ci}}[2\pi(k-i)]+(-1)^{k+i+1}\ln
\frac{1+k+i}{k-i}\Bigg\}.
\end{equation}
Diagonal elements have the form
\begin{equation}\label{so5}
\beta_{ii1}=\frac{1}{\pi}\Big\{-\gamma+{\mathrm {Ci}}[2\pi(1+2i)]-\ln [2\pi(1+2i)]\Big\},\ \gamma=0.577216...
\end{equation}
Here Ci$(x)$ is cosine integral function \cite{abr}. Analogously one proceeds with
 $\gamma_{ik1}$, which can be analytically  computed as  well,  \cite{mypre}.

For  the odd subspace we have
\begin{equation}\label{od1}
\sum_{k=0}^\infty b_k (\eta_{ki\alpha }+\zeta_{ki\alpha })=Eb_{i\alpha },
\end{equation}
where $\eta_{ki\alpha }$ is given by Eq. (35) of Ref. \cite{mypre} and
\begin{equation} \label{od2}
\zeta_{ik\alpha }=-\frac{A_\alpha }{\alpha } \int_{-1}^1 \sin k\pi x  \sin i\pi x\ \frac{(1-x)^\alpha +(1+x)^\alpha }{(1-x^2)^\alpha }dx.
\end{equation}
For concretness,  we reproduce a  computation  outcome   in the  special (Cauchy) case $\alpha =1$:
\begin{eqnarray}
\zeta_{ik1}=\frac{1}{\pi}\Bigg\{(-1)^{k-i}{\mathrm {Ci}}[2\pi(k+i)]+(-1)^{k+i+1}{\mathrm {Ci}}[2\pi(k-i)]+(-1)^{k+i+1}\ln
\frac{k+i}{k-i}\Bigg\},\label{od3} \\
\zeta_{ii1}=\frac{1}{\pi}\Big\{-\gamma+{\mathrm {Ci}}[4i\pi]-\ln [4i\pi]\Big\}.\label{od4}
\end{eqnarray}

For $\alpha  \neq 1$  the same arguments  are valid and can be safely reproduced step by step. The only difference is that matrix elements need to be  calculated numerically.  To obtain the results at the reasonable time cost, we  use smaller-sized   matrices,  around 50x50. This gives  access to two decimal places for (lowest) eigenvalues and a  fairly good approximation of the corresponding  eigenfunctions.

Six  lowest eigenvalues for the equation  \eqref{er3} are reported in  Table I,  for each of the  stability indices  $\alpha =0.5$, $1.0$ and $1.5$.  Quite good  coincidence is seen  with the spectral   data reported in Table 1 of Ref. \cite{duo} (obtained by an alternative fractional Laplacian discretization method).  Since our   main purpose has been to test the  computation method   of \cite{mypre0,mypre} against  an  alternative proposal of Refs. \cite{duo,duo1}, we refrain from a comparative  listing  of  other eigenvalues and other $\alpha $ choices, see  however  \cite{duo}.

\begin{table} [tbh]
	\begin{tabular}{|c|c|c|c|c|c|c|}
		\hline
		$i$ & 1 & 2 & 3 & 4 & 5 & 6 \\
		\hline
		$\alpha =0.5$ & 0.0048 & 0.4495 & 0.8724 & 1.2189 & 1.5041 & 1.7809 \\
		$\alpha =0.5$ \cite{duo} & 0.0038 & 0.4593 & 0.8626 & 1.2091 & 1.5149 & 1.7911\\
		\hline
		$\alpha =1.0$ & 0.1177 & 1.1926 & 2.5888 &4.0147 & 5.5328 & 7.0077 \\
		$\alpha =1.0$ \cite{duo} & 0.1135 & 1.2026 & 2.5760 & 4.0292 & 5.5171 & 7.0245\\
		\hline
		$\alpha =1.5$ & 0.8059 & 3.6475 & 7.7541 & 12.816 & 18.676 & 25.230 \\
		$\alpha =1.5$ \cite{duo} & 0.8088 & 3.6509 & 7.7500 & 12.811 & 18.670 & 25.235\\
		\hline
	\end{tabular}
	\caption{Table of 6 lowest eigenvalues of the regional fractional Laplacian for $\alpha=0.5$, 1 and 1.5 calculated for the 	 2000x2000  matrix, composed with repect to the trigonometric base  with Dirichlet boundary conditions. Energy levels are numbered according to standard  convention: the	lowest (ground) state is  labeled by
	$n=1$ (even state with $k=0$), $n=2$ corresponds to lowest odd state with $k=1$,
 $n=3$ corresponds to even state with $k=2$ etc. The corresponding eigenvalues
 from Ref. \cite{duo} are listed for comparison.} \label{shh}
\end{table}

\begin{figure}
\centering
\includegraphics[width=0.7\columnwidth]{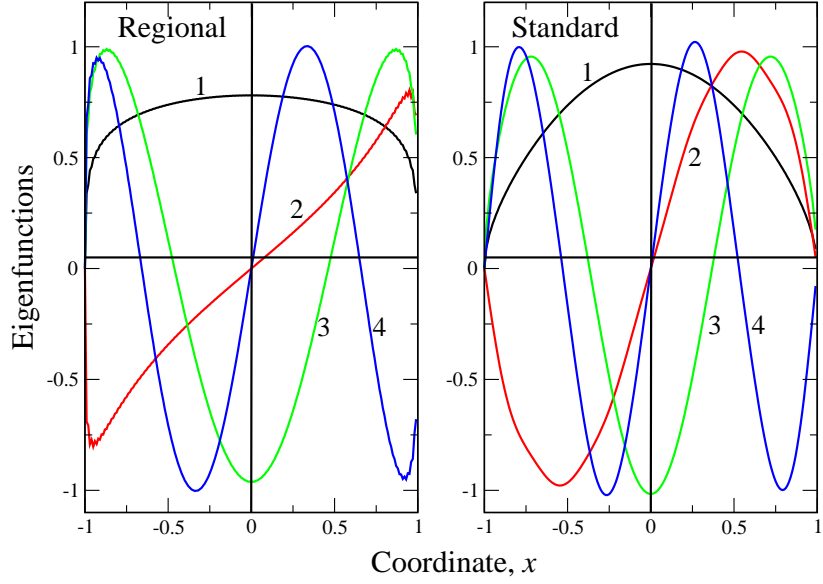}
\caption{Comparison between four lowest eigenfunctions of  the  regional (left panel) and restricted  (right panel) fractional Laplacians for $\alpha =1$. Eigenstate labels $n=1, 2, 3, 4$ are indicated.} \label{ig2}
\end{figure}

\begin{figure}
\centering
\includegraphics[width=0.7\columnwidth]{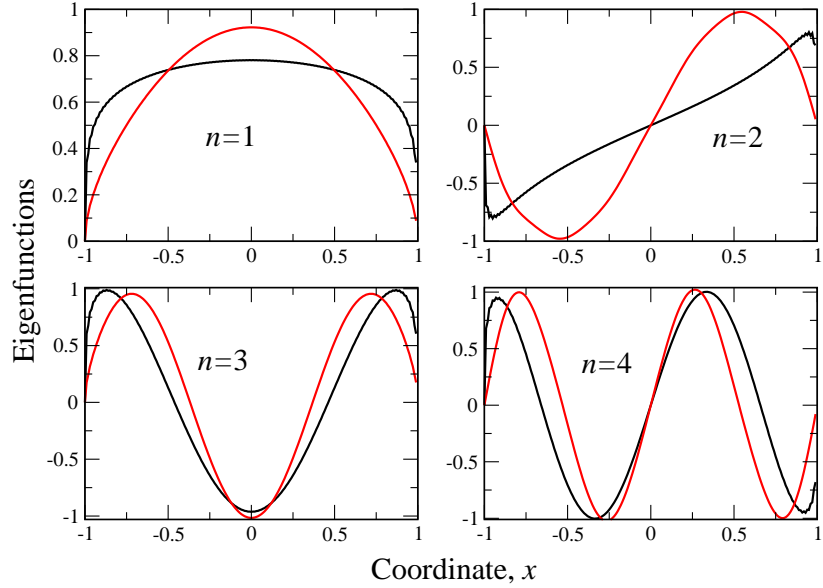}
\caption{Comparison of first four  eigenfunctions of regional and restricted fractional Laplacians, while   displayed  in pairs corresponding to $n=1, 2, 3, 4$. Black curves refer to the regional   Laplacian, red  curves to the restricted one; $\alpha =1$.} \label{ig3}
\end{figure}

Four lowest eigenfunctions for $\alpha =1$ are reported in Fig. \ref{ig2}. It is seen (compare e.g.  left and right panels of Fig. \ref{ig2}) that the eigenfunctions for regional and standard fractional Laplacians are qualitatively similar with one exception. Namely, those for regional operator show up much  sharper decay as $x \to \pm1$ and  their derivatives  diverge  to infinity as the boundary points are appproached.

More detailed comparative display of Fig. \ref{ig3}  gives further  support to   our statement    about the  sharp decay (steep decent down to zero) of  the  regional fractional Laplacian  eigenfunctions, while   set against these for  the  restricted  one.
Comparing Eqs. \eqref{ger5} with \eqref{er3}, we realize that the milder  decay  in the restricted case  is a consequence   of a strong repulsion (scattering) from the boundaries, that is encoded in the functional form of the   (inverted)  singular potential \eqref{er2}  in the eigenvalue problem of the form (13).

We note that our results are consistent with independent  findings of Ref. \cite{duo} (that refers as well to a different computation method).  C.f. Fig. 7  therein, where the first and second eigenfunctions of the restricted and regional fractional Laplacian were compared. Additionally these for the spectral fractional Laplacian have been depicted.

\begin{figure}
\centering
\includegraphics[width=0.7\columnwidth]{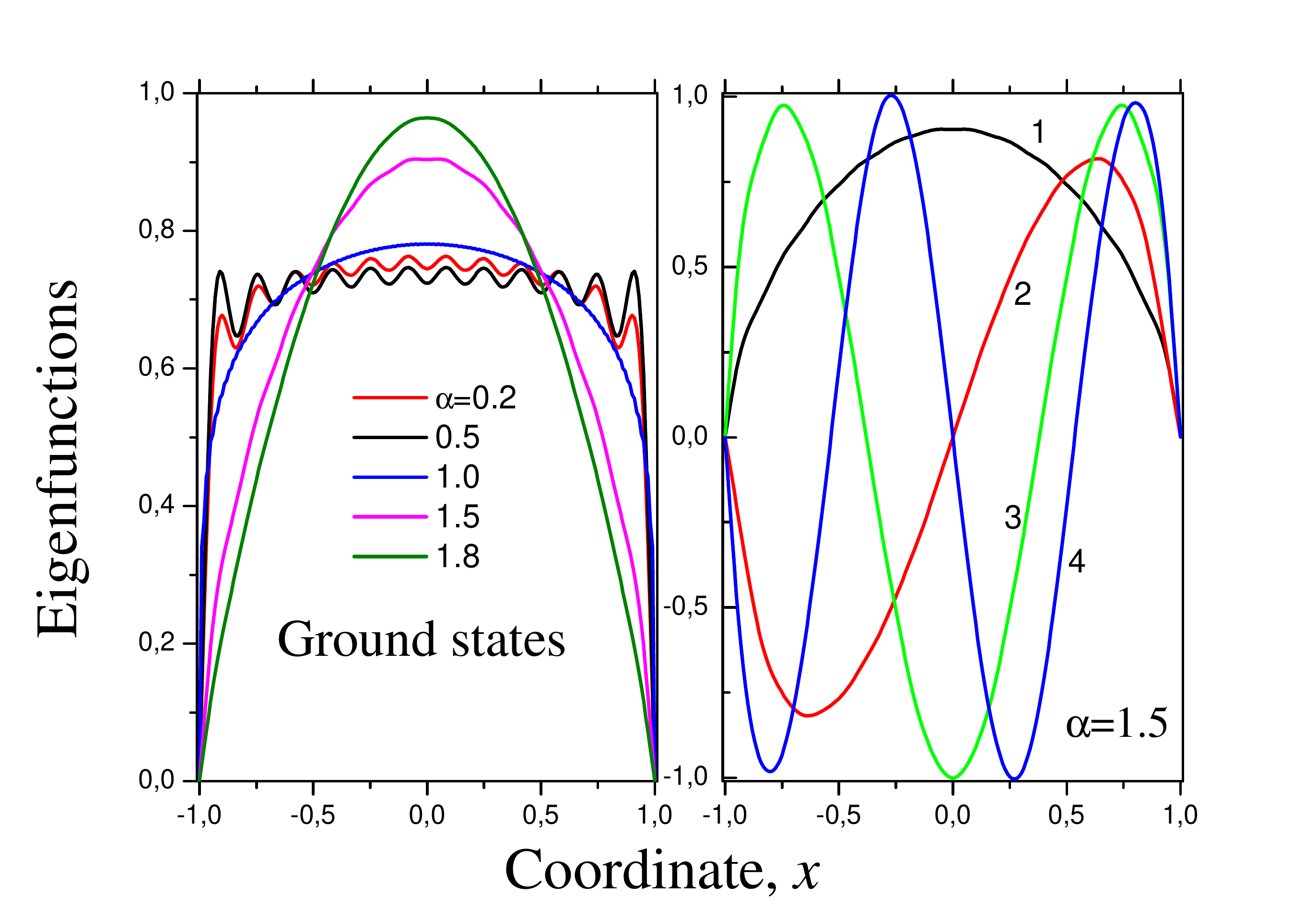}
\caption{Left panel. Comparison of ground states for the regional fractional operator \eqref{er1} for different L{\'e}vy indices $\alpha $, shown in the panel. Right panel portrays the same as in Fig. \ref{ig2} but for $\alpha =1.5$. Figures near curves correspond to the number of a state.} \label{ig4}
\end{figure}

To give  a glimpse of the $\alpha $-dependence of the discussed spectral problems, in Fig. \ref{ig4} we  display comparatively    the  $n=1$ eigenfunctions (ground states) for a couple of    $\alpha  \neq 1$  values  (the left panel).  We note that   for  $\alpha =0.2$ and $0.5$ the ground state eigenfunctions show up an  oscillatory behavior,  close to a maximum.  The computation has been completed for  50$\times$50  matrices  (and checked for 30$\times$30, which still can be viewed  to provide  too  rough approximation.   These oscillatory artifacts are expected to smoothen down with the matrix size growth.
  For $\alpha =1.5$ and 1.8, curves have been evaluated by employing   50$\times$50 sized matrices, and are smooth.

  The  right panel of  Fig. \ref{ig4} depicts four lowest   eigenfunctions  for   $\alpha =1.5$. The  pertinent   curves  are qualitatively similar to those for $\alpha =1$, c.f. Fig. 3.

\subsection{Regional fractional Laplacian. Trigonometric base with Neumann
 boundary conditions - an obstacle.}

 On the basis of reults reported in Figs. \eqref{ig2} to   \eqref{ig4}, it  is possible to investigate  the behavior of the  derivative of the  ground state in the vicinity of  boundary points. Clearly,   $\psi_1'(x\rightarrow \pm 1)$ becomes smaller as $\alpha $ increases. While for $0.2<\alpha <1$ the decay of $\psi_1(x)$ is steep, at $\alpha >1$ the decay becomes progressively milder. Accordingly,  as $\alpha $ increases, the derivative  $\psi_n'(x\rightarrow \pm 1)$ of any state decreases.  It  is different from zero in the whole  range $0<\alpha <2$.

 It is thus natural to address the question of whether the traditional Neumann condition (vanishing od the derivative of the function at the boundaries)  is at all feasible for the regional fractional Laplacian.   The natural choice  at this point is to pass from the Dirichlet to the Neumann $L^2([-1,1])$ basis (originally devised for the standard Laplacian in the interval, compare e.g. subsection III.B,  Eqs. (24) and (25)).

The pertinent Neumann base in the interval of length  $L$  comprises $f_n=\cos(n\pi x/L)$, $0<x<L$, $n=0,1,2,...$.   In the dimensionless units (measuring $x$ in the units of $L$) the functions assume the form  $f_n=\cos n\pi x,\ x \in [0,1],\ n=0,1,2,3,...$
This  basis  system  is orthogonal but not normalized: $
\int_0^1 \cos n\pi x \cos k \pi x dx= \frac 12 \delta_{nk},\ n,k \neq 0 $  or  $1,\quad  n=k=0$.
The passage to the  interval [-1,1] is accomplished via a  substitution $x \to (x+1)/2$, which gives rise to $f_n=\cos \frac{n\pi}{2}(x+1),\ n=0,1,2,...,\ x \in [-1,1]$.  This basis system is orthonormal in $L^2([-1,1])$ except for $f_0=1$.  After incorporating the normalization coefficient,  the orthonormal base with Neumann boundary conditions  at endpoints of $[-1,1])$   reads: $ \{
f_0=1/\sqrt{2}, \,  f_n=\cos \frac{n\pi}{2}(x+1),\ n=1,2,...,  x \in [-1,1]\}$.

\begin{figure}
\centering
\includegraphics[width=0.6\columnwidth]{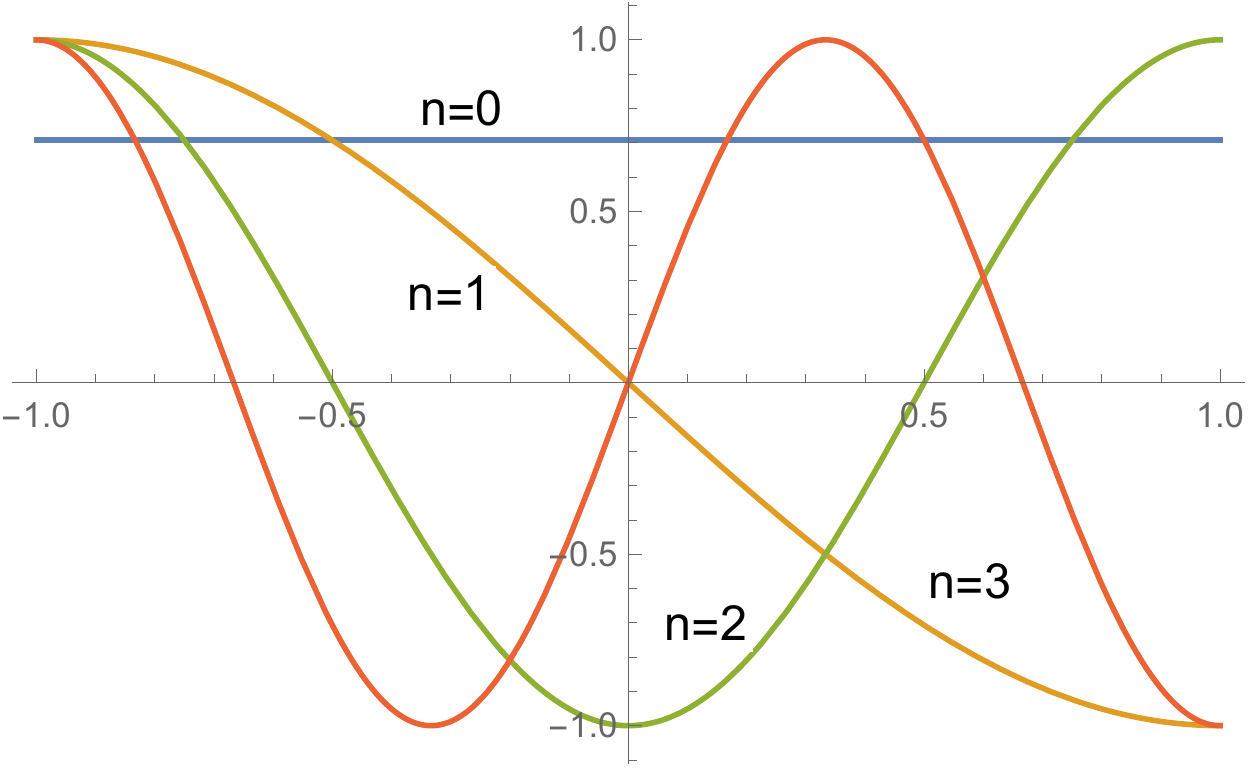}
\caption{Plot of four first functions of the  Neumann  base  in the interval $[-1,1]$.} \label{nnb}
\end{figure}

The Neumann basis  functions take non-zero  values at the boundaries $x= \pm1$, see e.g.  Fig. \ref{nnb},  and  this fact actually precludes the convergence of integrals that  define  matrix elements  involved in the solution of  the eigenvalue problem  for  the regional operator.

To demonstrate the divergence  obstacle, we calculate explicitly the auxiliary function $g_{k\alpha }$ (see Eq. (15) of Ref. \cite{mypre}) for $\alpha =1$. We have
\begin{eqnarray}
{\color{red}{g_{k1} }}&=&-\frac{1}{\pi}\int_{-1}^1\frac{\cos \frac{k\pi}{2}(x+1)}{(z-x)^2}dz=\frac{1}{\pi}\left(\frac{\cos k\pi}{1-x}+\frac{1}{1+x}\right)+\frac k2 \Bigg\{\left[{\mathrm {Ci}}\frac{k\pi}{2}|x-1|-{\mathrm {Ci}}\frac{k\pi}{2}|x+1|\right]\sin \frac{k\pi}{2}(x+1)+ \nonumber \\
&&+\left[{\mathrm {Si}}\frac{k\pi}{2}(x+1)-{\mathrm {Si}}\frac{k\pi}{2}(x-1)\right]\cos \frac{k\pi}{2}(x+1)\Bigg\}.\label{pam5}
\end{eqnarray}

Now, if we pass to an explicit form of the matrix elements  (c.f. Eq. 18) of Ref. \cite{mypre})
\begin{equation} \label{pam6}
\gamma_{kl1}=\int_{-1}^1 \cos \frac{l\pi}{2}(x+1) g_{k1}(x)  dx,
\end{equation}
we see that the terms in Ex. \eqref{pam5}, which   contain $(1\pm x)^{-1}$,  generate  the  logarithmic divergence.

The choice of  $\alpha =1$ may be considered special, but  explicit computations allow to identify jeopardies to be met, if  the Neumann base in use.  Actually, the situation is more  intricate. In below  we shall see that  for  $\alpha  \geq 1$  the divergence problem persists,  while for  $\alpha  < 1$    one can handle the integrals.

Interestingly, these observations  appear to   stay in conformity with the mathematically rigorous  discussion of divergence jeopardies in case  of censored L\'{e}vy  flights,  \cite{bogdan},  and  of reflected L\'{e}vy flights proper, \cite{warma}, where the stability parameter ranges $0<\alpha <1$ and $1\leq \alpha <2$  were found to   refer to qualitatively different jump-type processes.

 We now return to the above functions $f_n=\cos \frac{k\pi}{2}(x+1)$ ($k=0,1,2,...$) comprising complete orthonormal basis system in $L^2(D), D=[-1,1]$. While seeking a solution  $\psi _{\alpha }(x)$   of the eigenvalue problem for the   fractional Laplacian, we     consider the expansion $\psi _{\alpha }(x)  =   \sum_{k=0}^{\infty } a_{k\alpha }\, f_k(x)$.

 To  quantify  a possible outcome  of the choice of  Neumann boundary conditions,  it is sufficient to consider the first  integral  term in Eq. \eqref{er1},  which is  a remnant of the action of the  restricted  fractional Laplacian proper   upon $\psi (x)$.  Clearly,  the second term (the inverted potential)  does not depend on the choice of the (Dirichlet vs Neumann)  boundary conditions.

 Following the arguments of \cite{mypre}, we invoke    Eq. (15)  therein, and  note that for an auxiliary  function  $g_{k\alpha }(x)$ ($0<\alpha <2$) we actually have:
\begin{eqnarray}
g_{k\alpha }(x)=-A_\alpha \int_{-1}^1 \frac{ f_k(u)}{|u-x|^{1+\alpha }}=\{x-u=t,\ u=x-t,\ du=dt \}=-A_\alpha \int_{x-1}^{x+1}\frac{ f_k(x-t) dt}{t^{1+\alpha }}=\nonumber \\
=-A_\alpha  \left\{\begin{array}{cc}
\frac{dt}{t^{1+\alpha }}=dv & u=f_k(x-t) \\
v=-\frac{1}{\alpha  t^\alpha } & du=-f'_k(x-t)dt
\end{array} \right\}=
-A_\alpha \left[-\frac{f_k(x-t)}{\alpha  t^\alpha }\Biggl|_{x-1}^{x+1}-\frac{1}{\alpha }\int_{x-1}^{x+1}\frac{f'_k(x-t)dt}{t^\alpha } \right]=\nonumber \\
=\frac{A_\alpha }{\alpha }\left[\frac{ f_k(1)}{(x-1)^\alpha }-\frac{f_k(-1)}{(x+1)^\alpha }+\int_{x-1}^{x+1}\frac{f'_k(x-t)dt}{t^\alpha } \right].\label{am1}
\end{eqnarray}

Would we have imposed  the Dirichlet boundary conditions $f_k(\pm1)=0$, the first two terms in Eq. \eqref{am1} would  vanish, leaving us with convergent integral. In the non-Dirichlet regime the situation becomes more complicated, since we  may encounter divergent   matrix elements.  The latter obstacle we shall  analyze in more detail.

To this end   let us calculate  the matrix elements $\gamma_{\alpha  kl}= \int_{-1}^1 g_{k\alpha }(x) f_l(x)dx $ of the (so far) {\em{ordinary}}  fractional Laplacian:
\begin{equation}\label{am2}
\gamma_{\alpha  kl}=\frac{A_\alpha }{\alpha }\int_{-1}^1f_l(x)\left\{\frac{f_k(1)}{(x-1)^\alpha}-\frac{f_k(-1)}{(x+1)^\alpha} +\int_{x-1}^{x+1}\frac{f'_k(x-t)dt}{t^\alpha} \right\}dx.
\end{equation}
It can be shown that the double integral in Ex. \eqref{am2} is convergent (of course in the sense of Cauchy principal value) for all $0<\alpha <2$. At the same time if $f_k(\pm 1)\neq 0$, we should consider the first two integrals separately. We have
\begin{equation}\label{am3}
I_1=\int_{-1}^1\frac{ f_l(x)dx}{(x-1)^\alpha },\ I_{-1}=\int_{-1}^1\frac{ f_l(x)dx}{(x+1)^\alpha }.
\end{equation}
The "dangerous" points are $x=1$ for $I_1$ and $x=-1$ for $I_{-1}$.  Clearly, in the Dirichlet case  $q_l(\pm 1)=0$, no  convergence  problem  arises.

However, if $f_l(\pm 1) =const \equiv C$ (with respect to the convergence issue,  it does not matter  whether  these constants can or cannot be different at $x=1$ and $x=-1$), we have
\begin{eqnarray}\label{am4}
&&I_1(x \to 1)=C\int_{-1}^1\frac{dx}{(x-1)^\alpha }=C\frac{(x-1)^{1-\alpha }}{1-\alpha }\Bigg|_{-1}^1=\{x-1=\delta,\ x \to 1, \ \delta \to 0\}
\sim \frac{C}{1-\alpha }\lim_{\delta \to 0}\delta^{1-\alpha }=\nonumber \\
&&=\left\{ \begin{array}{c}
0, \qquad \quad \alpha <1 \\
\ln \delta \to \infty,\  \alpha =1 \\
\infty,\ 1<\alpha <2.
\end{array}  \right.
\end{eqnarray}
The same behavior is shared by  $I_{-1}$ near $x=-1$. This shows that regardless the value of derivative $f'_l(\pm 1)$, the integrals \eqref{am3} are divergent in the range $1\leq \alpha <2$.

W point out that a departure point for our discussion was the Neumann basis system and thus standard Neumann boundary data (vanishing of the derivative at the endpoints of $[-1,1]$) were implicit.  The outcome (51) tells us that   the regional fractional Laplacian may react consistently to Neumann boundary data in the range $0<\alpha <1$.

This observation stays in conformity with  results of mathematical papers  \cite{guan,warma}, where the range $0<\alpha <1$ has been singled out for the unquestionable identification of the regional fractional Laplacian on a closed bounded domain as  the generator of a reflected $\alpha $-stable   process.  Interestingly, no  traditional form of the Neumann condition has been in use.
On the other hand, a traditional looking (merely on the formal, notational level)  Neumann-type boundary condition   has been found to   be  necessary  for  the existence of the reflected process in the range $1\leq \alpha <1$.

\section{Regional fractional Laplacian: Signatures of reflecting boundaries.}

The explicit   spectral solution for the regional fractional Laplacian in the non-Dirichlet regime is not known in the  literature, except for some general existence  statements, \cite{bogdan,guan,warma, grubb}.   The low part of that spectrum (shapes of  ground and first excited eigenstates, related egenvalues)  remains unkonwn as well.  The spectral solution   reported in \cite{duo}  refers explicitly to the Dirichlet boudary data for the regional fractional Laplacian.

Our major purpose is to deduce the spectral solution that would have something in common with physicists' intuitions about reflected random motions in a bounded domain. To this end we shall  address  the spectral problem for the regional fractional Laplacian more carefully, avoiding  the decomposition of the  (nonlocal
operator action-defining)  integral expression into a sum of  integrals,
 of the form (27) or (31) -(34).  We shall  not impose any explicit form of the Neumann condition (or any of its analogs, that can be met in the literature, \cite{guan,warma,barles,dipierro,abtangelo1}).  The only Neumann input will  be related to the choice of the  basis system in $L^2(D)$, the latter Hilbert space  being  not  the one favoured by mathematicians \cite{guide}.  Lowest eigenvalues and shapes of related eigenfunctions  will be deduced with a numerical assistance.

The structure of expression for fractional regional operator \eqref{er1} shows that the term $\psi(x)$ balances $\psi(u)$ in the integrand numerator making the integral  convergent.   Indeed,  by inserting  $u=x+\delta$ in the integrand \eqref{er1} and expanding at small $\delta=u-x$ in power series, we obtain that around  the
dangerous point $u=x$, the integral takes  the form $\int\frac{u-x}{|u-x|^{1+\alpha }}du$, which is convergent for $0<\alpha <1$.
Note that it  has the  form $\int dt/t^\alpha  \sim t^{1-\alpha }=0$ at $\alpha <1$) and exists as  the  Cauchy principal value  for
 $1\leq \alpha  \leq 2$.    We have previously discussed this question in Ref.\cite{pre11}, see Eqs. (9), (10) and surrounding discussion therein. We shall
 analyze  that issue  in more detail, in  below.

Accordingly,  to calculate safely (i.e. without   divergencies) the spectrum of  the  regional operator \eqref{er1}, we should not split the integral into  a sum of terms  containing  respectively $\psi(u)$ and $\psi(x)$, but rather consider them together.
(We  point out that in case of the restricted fractional Laplacian with
 Dirichlet boundary data,  we were   actually   urged to split the integral into a sum, because the term with $\psi(x)$  has been vanishing identically in this case, see Ref. \cite{mypre} for details.)

Let  us make use of the modulus property
\begin{equation} \label{do1}
|u-x|=\left\{
\begin{array}{c}
u-x,\ u>x \\
x-u,\ u<x.
\end{array} \right\}
\end{equation}
and  rewrite the limits of integration in Eq. \eqref{er1} accordingly, so arriving at (once more here $D=[-1,1]$)
\begin{equation}\label{do2}
(-\Delta )^{\alpha /2}_{D,reg}\psi(x)=-A_\alpha  \int_{-1}^1 \frac{\psi(u)-\psi(x)}{| u-x|^{1+\alpha }}du=-A_\alpha  \left[\int_{-1}^x \frac{\psi(u)-\psi(x)}{(x-u)^{1+\alpha }}du+\int_x^1 \frac{\psi(u)-\psi(x)}{(u-x)^{1+\alpha }}du\right].
\end{equation}
We perform the substitution $x-u=t$ ($u=x-t$) in the first integral to obtain
\begin{equation}\label{do3}
P_1(x)=\int_0^{x+1}\frac{\psi(x-t)-\psi(x)}{t^{1+\alpha }}dt.
\end{equation}
Next we substitute  $u-x=t$ ($u=x+t$) in the second integral:
\begin{equation}\label{do4}
P_2(x)=\int_0^{1-x}\frac{\psi(x+t)-\psi(x)}{t^{1+\alpha }}dt.
\end{equation}
The  regional operator can be rewritten in the form
 \begin{equation}\label{do5}
(-\Delta )^{\alpha /2}_{D,reg}\psi(x)=-A_\alpha \bigg[P_1(x)+P_2(x)\bigg].
\end{equation}

We shall  demonstrate that  for $0< \alpha <1$ the integrals $P_i(x)$ ($i=1,2$) are convergent  as $t\to 0$.   We have in the lowest order in $t$
\begin{equation}\label{do6}
\psi(x-t)\approx \psi(x)-t\psi'(x),\ \psi(x+t)\approx \psi(x)+t\psi'(x).
\end{equation}
Substitution of \eqref{do6} into \eqref{do3} and \eqref{do4} yields
\begin{eqnarray}
P_1(x)=-\psi'(x)\int_0^{x+1}\frac{tdt}{t^{1+\alpha }}=-\psi'(x)\left.\frac{t^{1-\alpha }}{1-\alpha }\right|_0^{x+1}=-\psi'(x)\frac{(x+1)^{1-\alpha }}{1-\alpha },\ \alpha <1.\label{do7} \\
P_2(x)=\psi'(x)\int_0^{1-x}\frac{tdt}{t^{1+\alpha }}=\psi'(x)\left.\frac{t^{1-\alpha }}{1-\alpha }\right|_0^{1-x}=\psi'(x)\frac{(1-x)^{1-\alpha }}{1-\alpha },\ \alpha <1.\label{do8}
\end{eqnarray}
We note that at $1 \leq \alpha  \leq 2$ we should   interpret  $P_1+P_2$  as a  whole, but in terms of  the Cauchy principal value procedure.  The limiting  behavior near zero,  gives  rise to  the cancellation similar to  that in  Eq. (10) from Ref. \cite{pre11}.

In  the lowest order in $t$,  the final form of of the regional fractional operator
reads
\begin{equation}\label{do9}
(-\Delta )^{\alpha /2}_{D,reg} \psi(x)\approx -A_\alpha  \frac{\psi'(x)}{1-\alpha }\biggl[(1-x)^{1-\alpha }-(1+x)^{1-\alpha }\biggr].
\end{equation}
The expression \eqref{do9} is already free from  divergencies.

We note that  the procedure \eqref{do6} can be extended to an  arbitrary order in $t$ to yield the representation of the regional fractional Laplacian   (nonlocal, integral operator) through the infinite series of  locally defined  differential operators.  For our present purposes we find  that "series idea" is impractical, being too much computing time consuming. Therefore,  we shall compute explicitly  the integrals  \eqref{do2} substituting $\psi (x)$ by $f_k(x)$, which are the elements of the Neumann basis in $L^2(D)$, $n=0,1,2,...$ . We mind the structure of the integrands in Eqs \eqref{do2} - \eqref{do9}. Namely, the L\'evy index $\alpha$ is contained only in the denominators, while functions
$\psi (x)$ do not have this index. The same situation occurs for the functions $f_k(x)$. Below, for convenience, we include index $\alpha$ in the definition of $f_k(x)$, i.e. we put formally $f_k(x) \equiv f_{k \alpha}(x)$. We have:

\begin{eqnarray}
&&f_{k\alpha }(x)=-A_\alpha \bigg[P_{1,k\alpha }(x)+P_{2, k \alpha }(x)\bigg],\, \,  P_{1,k\alpha }(x)=\int_0^{x+1}\frac{\cos\frac{k \pi}{2}(x-t+1)-\cos\frac{k \pi}{2}(x+1)}{t^{1+\alpha }}dt= \nonumber \\
&&=\cos\frac{k \pi}{2}(x+1)P_{11,k\alpha}(x)+\sin\frac{k \pi}{2}(x+1)P_{12,k\alpha}(x). \label{re3} \\
&&P_{2,k\alpha }(x)=\int_0^{1-x}\frac{\cos\frac{k \pi}{2}(x+t+1)-\cos\frac{k \pi}{2}(x+1)}{t^{1+\alpha }}dt=\nonumber \\
&&=\cos\frac{k \pi}{2}(x+1)P_{21,k\alpha}(x)-\sin\frac{k \pi}{2}(x+1)P_{22,k\alpha}(x). \label{re4} \\
&&P_{11,k\alpha}(x)=\int_0^{x+1}\frac{\cos \frac{k\pi t}{2}-1}{t^{1+\alpha }}dt,\ P_{12,k\alpha}(x)=\int_0^{x+1}\frac{\sin \frac{k\pi t}{2}}{t^{1+\alpha }}dt, \label{re5a} \\
&&P_{21,k\alpha}(x)=\int_0^{1-x}\frac{\cos \frac{k\pi t}{2}-1}{t^{1+\alpha }}dt,\ P_{22,k\alpha}(x)=\int_0^{1-x}\frac{\sin \frac{k\pi t}{2}}{t^{1+\alpha }}dt.
\label{re5}
\end{eqnarray}

To derive the expressions for  $P_{ij,k\alpha}$ ($i,j=1,2$), we use following trigonometric identity: $\cos(A-B)=\cos A\cos B + \sin A \sin B$ so that for example in \eqref{re3}
\begin{eqnarray*}
\cos\frac{k \pi}{2}(x+1-t)-\cos\frac{k \pi}{2}(x+1)=\cos\frac{k \pi}{2}(x+1)\cos\frac{k \pi t}{2}+\sin\frac{k \pi}{2}(x+1)\sin\frac{k \pi t}{2}-\cos\frac{k \pi}{2}(x+1)=\\
=\cos\frac{k \pi}{2}(x+1)\left(\cos\frac{k \pi t}{2}-1\right)+\sin\frac{k \pi}{2}(x+1)\sin\frac{k \pi t}{2}.
\end{eqnarray*}
The last expression, being divided by $t^{1+\alpha}$ and integrated, yields Eq. \eqref{re5a} for $P_{11,k\alpha}(x)$ and $P_{12,k\alpha}(x)$. The derivation of Eqs \eqref{re5} is the same. In other words, the second subscript in the functions $P_{ij,k\alpha}$ ($i,j=1,2$) appears simply because the initial functions $P_{i, k \alpha }(x)$ ($i=1,2$) \eqref{re3}, \eqref{re4} contain two terms, each of which adds one more index $j=1,2$.

The integrals $P_{ij,k\alpha}$ ($i,j=1,2$) will be calculated numerically. For reference purposes we list the integrals $P_{ij,k\alpha}$ ($i,j=1,2$) in terms of variable $z=k\pi t/2$, which we use for actual numerical calculations
\begin{eqnarray}
P_{11,k\alpha}(x)=\left(\frac{k\pi}{2}\right)^\alpha  \int_0^{\frac{k\pi}{2}(x+1)} \frac{\cos z -1}{z^{1+\alpha }}dz,\ P_{12,k\alpha}(x)=\left(\frac{k\pi}{2}\right)^\alpha  \int_0^{\frac{k\pi}{2}(x+1)} \frac{\sin z}{z^{1+\alpha }}dz,\nonumber \\
P_{21,k\alpha}(x)=\left(\frac{k\pi}{2}\right)^\alpha  \int_0^{\frac{k\pi}{2}(1-x)} \frac{\cos z -1}{z^{1+\alpha }}dz,\ P_{22,k\alpha}(x)=\left(\frac{k\pi}{2}\right)^\alpha  \int_0^{\frac{k\pi}{2}(1-x)} \frac{\sin z}{z^{1+\alpha }}dz. \label{re6}
\end{eqnarray}
Also, to remove the (spurious) divergencies at $z=0$ "gracefully", we render the integrals to the form
\begin{equation}\label{re7}
P_{ij,k\alpha}(x)=\int_0^\delta +\int_\delta^{1\pm x}\equiv P_{ij,k0\alpha}(x)+\Delta P_{ij,k\alpha},\ \delta <0.01.
\end{equation}
The explicit expressions for $\Delta P_{ij,k\alpha}$ read
\begin{equation}
\Delta P_{11,k\alpha}=\Delta P_{21,k\alpha}=\left(\frac{k\pi}{2}\right)^\alpha  \frac 12 \frac{\delta^{2-\alpha }}{2-\alpha },\
\Delta P_{12,k\alpha}=\Delta P_{22,k\alpha}=\left(\frac{k\pi}{2}\right)^\alpha  \frac{\delta^{1-\alpha }}{1-\alpha }.
\end{equation}

Finally, the elements of the matrix $(\gamma_{kl\alpha })$, to  be diagonalized in order to find the desired spectrum,  are as follows
\begin{eqnarray}
\gamma_{kl \alpha }=-A_\alpha \int_{-1}^1 \cos\frac{l \pi}{2}(x-t+1)\Bigg\{\cos\frac{k \pi}{2}(x-t+1)\bigg[P_{11,k\alpha}(x)+P_{21,k\alpha}(x)\bigg]+ \nonumber \\
+\sin\frac{k \pi}{2}(x-t+1)\bigg[P_{12,k\alpha}(x)-P_{22,k\alpha}(x)\bigg]\Bigg\}dx. \label{gu}
\end{eqnarray}

The results of numerical calculations of the spectrum   for $(\gamma_{kl\alpha })$,   \eqref{gu},  are reported in the Table \ref{sh}. As computational times are very long (around six hours for 20x20 matrix and around 24 hours for 30x30 one), we limit ourselves for 20x20 matrices. However, the qualitative features of the spectrum remain the same as those for larger matrices. We have checked that by    test calculations of some of the  eigenfunctions, by means of 30x30 matrices.

\begin{table} [tbh]
\begin{tabular}{|c|c|c|c|c|c|c|}
\hline
$i$ & 1 & 2 & 3 & 4 & 5 & 6 \\
\hline
$\alpha =0.2$ & 0.0000 & 0.1871 & 0.3075 & 0.3972 & 0.4696 & 0.5307 \\
$\alpha =0.2$ \cite{duo} & 0.0003 & 0.1878 & 0.3085 & 0.3981 & 0.4700 & 0.5306\\
\hline
$\alpha =0.5$ & 0.001 & 0.4499 & 0.8505 & 1.1959 & 1.5018 & 1.7785 \\
$\alpha =0.5$ \cite{duo} & 0.0038 & 0.4593 & 0.8626 & 1.2091 & 1.5149 & 1.7911\\
\hline
$\alpha =0.7$ & 0.004 & 0.6319 & 1.3138 & 1.9591 & 2.5665 & 3.1420 \\
$\alpha =0.7$ \cite{duo} & 0.0170 & 0.6729 & 1.3646 & 2.0140 & 2.6231 & 3.1993\\
\hline
$\alpha =0.9$ & 0.006 & 0.8290 & 1.8987 & 3.0064 & 4.1148 & 5.2156 \\
$\alpha =0.9$ \cite{duo} & 0.0640 & 0.9799 & 2.0823 & 3.2054 & 4.3230 & 5.4300\\
\hline
\end{tabular}
\caption{The comparison of 6 lowest eigenvalues $E_i$ of the regional fractional Laplacian, evaluated in the Neumann base,  for different $\alpha $  in the range $0<\alpha <1$  (calculated by means of the  20x20 matrices),   with those   obtained by imposing  the Dirichlet boundary condition. Data are   taken form Table I in Section IV of the present paper   and from   Table I  in  Ref. \cite{duo} (for the comparison purpose we  have  abstained from using the label zero; the displayed data should be read as first, second, third... eigenvalue of the regional Laplacian in question).   The first left column,   modulo the computing inaccuracy, is a  clear  signature of  reflecting boundary data that were imported  by executing the  the diagonalization in the Neumann basis. The ground state function in the  Neumann    base  is constant and the respective eigenvalue  should equal zero.  Our data show that numerically computed  lowest eigenvalue should  actually    be identified with zero,  up  to inaccuracies appearing in the  third decimal place.} \label{sh}
\end{table}

It seen that the Neumann base, in view of our matrix size limitations (small matrices, to lower the computing time)  gives  slightly inaccurate estimation for the lowest (ground state) eigenvalue . This is probably due to the fact that the lowest function of the   Neumann   base  $f_0=1/\sqrt{2}$ is simply constant.

\begin{figure}[tbh]
\centering
\includegraphics[width=0.7\columnwidth]{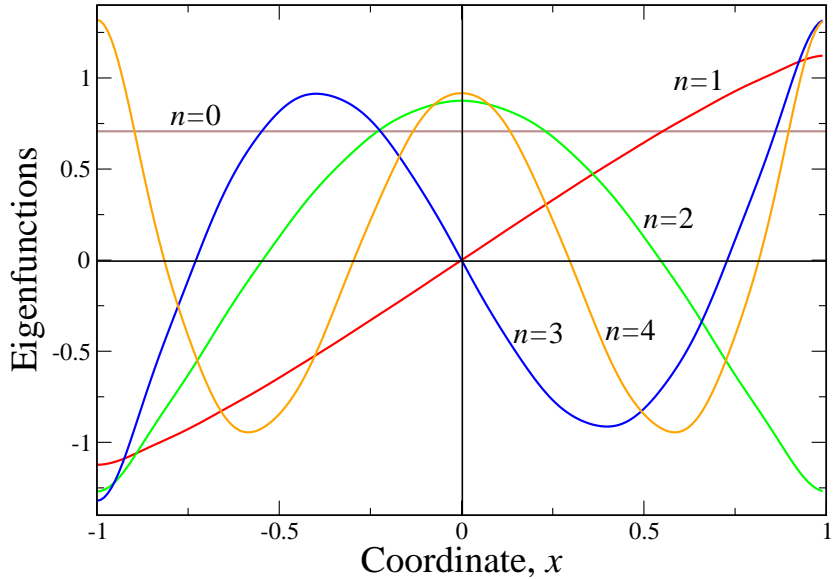}
\caption{The representative ($\alpha =0.5$) plot of the  first   five  eigenfunctions of the regional fractional Laplacian, found via the diagonalization in the Neumann base; $n=0$ function is constant,  being a clear signature of the reflecting boundary conditions.   The corresponding eigenvalue is  fapp (for all practical purposes) zero.  It is instructive to compare the shapes of eigenfunctions with  the elements of the Neumann base,  depicted in Fig. 5.} \label{sta}
\end{figure}

To convince ourselves that the obtained eigenvalues are signatures of reflecting boundaries, we have computed a couple of (approximate) eigenfunctions.
The representative plot of four lowest eigenfunctions for $\alpha =0.5$ is portrayed in Fig. \ref{sta}.

 We have checked  that for other $\alpha  \in (0,1)$  the eigenfunctions become   fairly  close to those  found for $\alpha =0.5$.   Also, they  are  not distant  from the  Neumann  basis  (trigonometric)  functions. Strictly speaking, up to a sign of resulting eigenfunctions $\psi_n$ may  happen to   be opposite to that of the Neumann trigonometric ones.  Except for the ground state, the sign issue is immaterial,
 sine it is  $|\psi_n|^2$  which   stands for the probability density.

The main feature of the eigenfunctions is that appear to  satisfy the Neumann boundary conditions, intepreted in terms of standard vanishing derivatives.

We have paid special attention to the range $0<\alpha <1$, because in this parameter regime   the integrals (53) are convergent.   The case of  $1< \alpha<2$  needs more attention, since these integrals exist only in the sense of Cauchy principal value.
 That enforces slight modifcations of the  matrix diagonalization procedure, previously  adopted  to calculate the spectrum  of   the  regional fractional Laplacian in the Neumann  base .

On one hand, the calculation of the matrix elements (integrals) is now more time consuming, beacuse we need to bypass the singularity at $u=x$ by means of   the  Cauchy principal value. On the other hand, the convergence of the matrix
method is generally much worse for the base with Neumann boundary conditions than that for Dirichlet ones.

 For instance,  for the Dirichlet base,   the  acceptable  accuracy  of computation outcomes has been achieved  already  for  20x20 matrices.  For  the Neumann base we need matrices   of the size 1000x1000   to achieve an acceptable accuracy  level  (the least  possible case is 600x600). That  substantially  increases the computational time.
The integration procedure for $1< \alpha<2$  incurs further increase of the computation time.

 By these reasons, we have explicitly checked the signatures of reflection by an explicit computation of  the ground state eigenvalue for $\alpha=1.2$, 1.5 and 1.8, followed by a computation of the shape of the corresponding eigenfunction.  As expected, the eigenvalues come up (for all practical purposes) as zero, and eigenfunctions are indeed constant.

In passing, we note, that the integrals (53) are identically zero for the constant function $\psi(x)$. This is actually the signature of  reflection, i.e.  the  zeroth ground state eigenvalue. \\

\section{On impenetrable barriers for L\'{e}vy flights: comments on  the stochastic behavior in  the vicinity of the boundary.}

The notion of censored stable processes, as introduced in Ref. \cite{bogdan} is
verbally rather loose: a censored stable process in an open set $D\subset R^n$ is obtained   by suppressing its jumps from D to the complement $R^n\backslash D$. Alternatively,  it  is a process "forced" to stay inside $D$.  Jumps are censored, i.e. these that  would  (according to the  jump-size  probability  law) "land" beyond $D$ are simply cancelled.  Next the process is resurrected at the stopping point and started anew. In connection with the  resurrection of Markov process se e.g.   \cite{bogdan,pakes}.

{\bf Remark 10:} The above concept  of resurrection, exploits the "starting  anew" property for the stopped  stochastic process. It shows some  affinity with the family of   stochastic   processes with   resetting, where   the wandering particle can be reset to an initial location, at a certain rate,  and  next the process is   started anew.  It is known that in the diffusion with stochastic resetting  one arrives at  nonequilibrium stationary states with non-Gaussian fluctuations for the  particle position. Apart form the above (reset) affinity, we have not found any  "probability accumulation"    outcomes that would resemble those reported in \cite{dybiec,dybiec1} and in the fractional Brownian motion on the interval, \cite{metzler2}.\\

The Authors of Ref. \cite{bogdan} exclude from considerations so-called  taboo processes  which are related to the concept of the Doob $h$-transform, \cite{gar1,mazzolo,pinsky} and  are known not to leave  the  open set $D$. It is demonstrated that  what  is   named in Ref. \cite{bogdan}  a censored processes  (actually,  a recurrent censored symmetric $\alpha $-stable process) is different (in law)  from the  conditioned not to leave $D$  symmetric stable one, c.f. pp. 104-105 in Ref. \cite{bogdan}.  That is by no mean a no-go statement for  taboo processes, quite simply  they do not involve any point-wise  censoring mechanism, since it is the conditioning that does the job (of not reaching the boundaries).

It is quite clear, that the above  loose definition encompasses both taboo and censored processes, plus (upon admitting that the boundary   of $D$ can be reached by the process)  $\alpha $-stable versions of reflected processes.
The main issue addressed in the mathematical literature has been to strengthen the concept of "reflection" by  inventing  nonlocal analogs of Neumann boundary conditions, \cite{guan,barles,warma,dipierro,abtangelo1}.

 Various scenarios of the behavior of the censored process in the vicinity of the boundary have been formulated  as well.  One  of them has been verbalized as  follows, \cite{dipierro,abtangelo1}: when the process  exits $D$, it  immediately comes back to $D$.  The way it comes back is:  if a process exits to a point  $x\in R^n\backslash \overline{D}$,  its return to $y\in D$ is realized with a probability density being proportional to $|x-y|^{-n-\alpha }$, hence not necessarily to the stopping point.

 A concept of the  resurrection for a Markov process has been invoked here as well, \cite{bogdan,pakes}.  It amounts to an   immediate resurrection  of the process  after eliminating  (censorship) the inadmissible jump (form  $D$  to $R^n\backslash D$),  through a procedure of gluing together stable processes:  at a stopping  point (and time) of the  tentativelyy  terminated process, we glue its copy that actually  gives birth to a process started  anew at the  stopping point.  The process proceeds  proceed up to the next  stopping time (i.e censored jump),  and the gluing procedure is repeated.   Such  continually  resurrected process is bound not to leave $D$, as required.

Leaving aside a great amount of technicalities  concerning
 the proper mathematical formulation of what a reflected stable process should  actually  be, we shall pass to a brief discussion of physicists'  viewpoint    on impenetrable barriers and eventually  on  the concept of reflection  (e.g. that of reflecting boundaries),  \cite{gitterman,buldyrev}, \cite{dybiec}-\cite{denisov}.

 Motivated by the path-wise simulations, physicists coin  their own  recipies  on how to implement the condition of reflection from the barrier in terms of the  the sample path  behavior (that on the computer simulation level).  For example,  in Ref. \cite{dybiec} the condition of reflection is assured  by wrapping the  trajectory,  destined to  hit the barrier (or crossing the barrier),  around the hitting point   location, while preseving the assigned length. On  the other hand, it is mentioned that for jump-type processes the location boundary is not hit by the majority of discontinuous sample paths  and returns (or recrossings the boundary location) should be excluded from considerations, which excludes the wrapping scenario form further considerations.

 Another viewpoint mentioned in Ref. \cite{dybiec}  refers to a simulation of the reflecting boundary by   an infinitely high hard (e.g. impenetrable) wall, quantum mechanically interpreted as  the  infite well  (\cite{zaba0,zaba1}),   yielding an immediate reflection once  its vicinity (and not necessarily the boundary itself) is  reached.

On the other hand, in Ref. \cite{dybiec1} another  proposal for  the   introduction of  the  reflection  scenario has been outlined (an idea inspired by \cite{broeck,denisov,dubkov}), with a focus on a numerical simulation of sample trajectories of the Langevin-type L\'{e}vy - stable evolution  in the binding extremally anharmonic potential $V_n(x)\sim x^n$ with $n  \gg 1$   (set  e.g. $n=800$ for concretness).

Here a  departure point  is an observation (strictly speaking, questionable in the quantum setting)  that in the limit $n\rightarrow \infty $  the potential $V_n(x)$ mimics the infnite well enclosure, with boundaries  at endpoints of the interval $[-1,1]$  (interpreted in \cite{dybiec1}  as reflecting).
The random motion is  described in terms of  the  Langevin-type  equation $dX(t)/dt= - {V'}_n(x) + \zeta _{\alpha }(t)$ where  $  \zeta _{\alpha }(t)$ is  a    formal encoding of the  symmetric white $\alpha $-stable noise, c.f. \cite{dybiec1}.
The limiting stationary   probability density of the associated fractional Fokker-Planck equation has been found \cite{denisov} in the form  of the $L^1([-1,1]$  normalized  function :
\be \label{eg1}
\rho _{\alpha } (x)= (2)^{1-\alpha }{\frac{\Gamma (\alpha )}{\Gamma ^2 (\alpha /2)}}
(1-x^2)^{\alpha /2-1},
\ee
the result  valid for  all  $0<\alpha \leq 2$.
The special case of the Cauchy noise ($\alpha =1$) has been addressed in Ref. \cite{dubkov},   by an
 independent reasoning,   with the outcome:
\be \label{eg2}
\rho _{1} (x) = {\frac{1}{\pi }} {\frac{1}{\sqrt{1 -x^2}}}
\ee
valid for all $|x| < 1$.  This function blows up to infinity at the boundaries of the interval $[-1,1]$.

In passing we note that for $\alpha =2$ a uniform Brownian distribution $1/L$ arises. That would suggest a link with reflected processes, but this observation is misleading.

The probability  density  functions  \eqref{eg1}, \eqref{eg2}, do not belong to the $L^2([-1,1])$  inventory associated with the regional fractional Laplacian. More than that, they refer to ranom motion scenarios that cannot reach the boundary and definitely comply with the stopping scenarios used in the numerical simulations in Ref. \cite{dybiec1}.
That  view is supported by the fact that \eqref{eg2} coincides with the familiar probability distribution function for the classical harmonic oscillator. Its high probability areas correspond to the long residence time for a classical particle in harmonic motion.

From the physical point of view,  the  most interesting observation of Ref. \cite{dybiec1} in this context is that the extremely anharmonic and stopping motion scenarios in the presence of the symmetric $\alpha $-stable noise, actually yield the same statistics of  simulation outcomes, c.f., Section II.A.2 in \cite{dybiec1}. Moreover, a   detectable  deviation from these outcomes has been reported if the wrapping scenario is employed.  Actually,  irrespective of the stability index $\alpha $,   the wrapping assumption, in the path-wise simulation procedure of Refs.  \cite{dybiec,dybiec1}   has been found to  lead to a uniform    asymptotic  distribution in the interval,  like in the reflecting Brownian motion.  The pertinent figures  have not been reporoduced in Ref. \cite{dybiec1}, but are available from the Authors, \cite{dybiec2}.

At this point we shall verbalize the concept of the stopping scenario, \cite{dybiec2}.  In the course of the path-wise simulation of the jump-type process, any jump longer than the distance of the point of origin from the boundary is cancelled (the process is stopped). In the $\varepsilon $-vicinity of the boundary, $\epsilon \ll 1$ all jumps in the direction of the barrier are cancelled (it is an explcit censorship at work). The process is kept stopped until the probability law will produce a jump in the direction opposite to the boundary, and next continued.

From a mathematical point of view it is clear that probability functions \eqref{eg1}, \eqref{eg2} are not eigenfunctions  of the regional fractional Laplacian.   Actually, after completing them in $R\backslash (-1,1)$  by assigning  them the value zero  beyond $(-1,1)$,  we realize that $\rho _{\alpha } (x)$ is the so-called $\alpha $-harmonic eigenfunction of the  original   fractional Laplacian   $(- \Delta )^{\alpha /2}$, Eq. (1),  defined on the whole of $R^1$ and     associated with  the eigenvalue zero:
\be \label{eg3}
(- \Delta )^{\alpha /2} \rho _{\alpha } (x) =0
\ee
c.f. \cite{what,kulczycki}.  One  should keep in mind that  the above idenity needs somewhat involved calculations in case of  arbitrary $1<\alpha<2$, but can be straightforwardly checked in case of $\alpha =1$. The calculation exploits in full a nonlocality of the fractional Laplacian, and terms that account for $R\{(-1,1)$  cannot be disregarded, \cite{dyda,private}.

 Thus a stochastic  process that is consistent with the pdf \eqref{eg2} is not the reflected one, but rather the censored one, see e.g. also \cite{bogdan}.
The pertinent  censored  process  never crosses or reaches  the boundary, which is a property shared with taboo processes  in the  impenetrable enclosure,  c.f.   the    fractional  infnite square well spectral problem  or   the related  taboo process in the interval  (so-called ground state process), \cite{zaba0,zaba1,mypre0,mypre,duo,duo1,gar}. Nonetheless, the associated  stationary probability distributions appear to  be very  diferrent,  behaving reciprocally   at the boundary:   quick decay  with a  lowering  distance from the boundary $ \sim (dist)^{\alpha /2}$  (taboo process),   versus blow up to infinity   $\sim 1/(dist)^{\alpha /2 -1}$  in the same regime  (censored process).  Clearly, the taboo case refers to a probability depletion  in the vicinity of the boundary, while its accumulation  close to the boundary   characterizes the considered   censored process.

\section{Outlook and prospects: Some advantages of the spectral lore for fractional Laplacians in bounded domains.}

\subsection{General considerations}

We have paid   special attention to spectral problems involved with nonlocal motion generators  (fractional Laplacians of arbitrary L\'evy index $0<\alpha < 2$), whose adjustment to account for  finite  (bounded)  spatial geometries is a source of ambiguities in the physical literature. Like in case of standard
diffusion processes, lowest eigenfunctions and  eigenvalues of the pertinent operators, are decisive  for  deducing measurable properties  of relaxation processes towards equailibrium or their near-equilibrium behavior.

Our  motivations  stem  basically  from a simply looking inquiry into the problem of all admissible stochastic processes in a bounded domain, that may be consistently associated (derived or inferred from) with  the primordial  L\'{e}vy  noise. That sets the  broad conceptual  context of  L\'{e}vy flights in  bounded domains  and
directly involves  a delicvate issue of properly defined boundary data for fractional Laplacians in finite geometries.

To this end, we have  adopted the numerically assisted  approach to the eigenvalue  problem of  fractional Laplacians in the interval,  developed by us earlier \cite{mypre0} and based on the expansion of the sought-for eigenfunctions  in the properly tailored orthonormal base of $L^2(D)$.   In the present paper,  we have  considered two such bases, namely trigonometric ones associated with $D=[-1,1]$.

The first one is familiar from the standard  quantum mechanical infinite well problem  \cite{land3}  and  comprises  functions with Dirichlet boundary conditions, i.e.  vanishing at  the boundaries. The second one is that obeying the Neumann boundary conditions,  i.e. basic functions are allowed to take arbitrary non-zero values  at the boundaries, while their derivatives are required to vanish. In both cases the spectral fractional Laplacian  (see Section II) simply imports all basic properties of the standard Laplacian. Things become complicated if other definitions of the fractional Laplacian in $D$ are considered.

In the course of the study, we  have demonstrated that the Dirichlet  base secures much better convergence properties of the numerical procedure, than the Neumann one. The latter base implies particularly bad convergence  properties in the stability parameter range  $1<\alpha <2$, where  integrals   of importance  (see, e.g., Eq. \eqref{do2}) exist only  in the sense of  Cauchy principal values. Not incidentally, in all studies devoted to censored stochastic processes, \cite{bogdan,guan},   the range $0< \alpha \leq 1$ has  been   considered separately from that of $1< \alpha < 2$, in  which major difficulties   were encountered, while attempting to give meaning to the notion of reflected jump-type processes and to devise a consistent form of Neumann boundary data.

We note in passing that in so-called  fractional quantum mechanics \cite{laskin}, the range $1 \leq \alpha < 2$  is introduced a priori as the one in which this theory is supposed  to make sense (it is not quite necessary  assumption  \cite{stef} that essentially narrows  the framework). We point out that a good reason for that may be the nonexistence of probability currents out of the  pertinent range \cite{stef}.  Notwithstanding,  consistent   spectral solutions  for Laplacians  in bounded  Dirichlet  domains are known to exist in the whole range $0<\alpha <2$).  On the other hand, spectral problems beyond  the range  $0<\alpha <1$,   make the Dirichlet boundary data the only  (for all practical purposes)   reliable choice,  against the Neumann  or Robin  ones), by purely pragmatic  reasons. That conforms with the standard quantum mechanical approach, where Neumann or Robin boundary data are definitely out of favor, while the Dirichlet data are prevalent.

To the contrary, all these data are often encountered and amply discussed in the theory of Brownain motion and diffusion-type processes. In connection with the eigenvalue problems addressed, we  note that one may in principle invoke the  variational approach, which is customary in the ordinary quantum mechanics. Such approach works for self-adjoint operators (see, e.g. \cite{bershub}) and hence should be valid for trial functions both with Dirichlet and Neumann boundary conditions  in more general than the standard Laplacian settings, i.e.  for fractional Laplacians as well. It would be  interesting to study the accuracy of the variational method for our two classes of  (trigonometric) trial functions.

\subsection{Fractional spectral problems: Exemplary association with  realistic  physical systems.}

Our discussion of section III on relevance versus irrelevance of eigenvalue problems  for fractional Laplacians, pertains to an interplay (intertwine) between the theory of stochastic processes  (stochastic modeling)  in a bounded domain and  the   fractionally generalized   quantum  model-systems (that refers to so-called fractional quantum mechanics, \cite{laskin0,laskin1,laskin}, see also \cite{stef})  which are spatially confined.    Leaving aside an explicit probabilistic (stochastic  processes)  viewpoint,  the developed   formalism can be applied to  physical systems of  confined geometry (like quantum wells and/or surfaces or interfaces), where the disorder and other kinds of   intrinsic  randomness (hence the fractional L\'{e}y noise)  can be interpreted on the quantum level.

 One class of such systems is an electronic ensemble, which tunnels through potential barriers in so-called spintronic devices (see \cite{zu,gvi,sar} and references therein). One of the realistic examples here is heterostructures like the LaAlO$_3$/SrTiO$_3$ interface \cite{oh,sd1,sd2}. To describe the experimental data related to above tunneling statistics,  fractional derivatives should be introduced in the conventional quantum-mechanical problem of a tunneling particle.

To be specific, the fractional generalization of the Schr\"odinger equation, describing the electronic properties of a hetero-junction between materials 1 and 2, reads (see Ref. \cite{sd1} for details)
\begin{equation}\label{ax1}
\mathcal{H}\Psi(z)\equiv\left( \begin{array}{cc}
s_c(z)+U(z) & iv(-\Delta)^{1/2} \\
iv(-\Delta)^{1/2} & -s_v(z)+U(z)
\end{array} \right) \left(\begin{array}{c}
\varphi(z)\\ \chi(z) \end{array} \right)=\varepsilon \left(\begin{array}{c}
\varphi(z)\\ \chi(z) \end{array} \right),
\end{equation}
where $\varepsilon$ is eigenenergy and $\Psi(z)$ is the spinor wave function of the electron at the interface. Here, $(-\Delta)^{1/2}$ is the fractional Laplacian for L\'evy index $\alpha=1$, $z$ is the dimensionless coordinate along the interface (normalized by the interface width $a$) \cite{sd1} and we use atomic units $\hbar=c=1$. The spatial confinement of the problem lie in the interfacial potential $U(z)$
\begin{eqnarray}
U(z)=U_0\delta(z)+U_1P(z), \label{ax2} \\
P(z)=\left[\theta(z)+\theta(d-z)\right],\nonumber
\end{eqnarray}
where the first term, containing $\delta$ - function, signifies the potential barrier exactly at the interface, while the second term denotes the barrier of height $U_1$ and width $d$, through which the electrons can tunnel. Here $\theta(z)$ is the Heaviside (unit step) function. The influence of technologically unavoidable disorder is modeled by the introduction of  fractional derivative $(-\Delta)^{1/2}$ instead of ordinary gradient $d/dz$. Our further notations are as follows \cite{sd1}
\begin{eqnarray}
	s_c(z)=s_{c1}[1-\theta (z)]+s_{c2}\theta (z), \nonumber \\
	s_v(z)=s_{v1} [1-\theta (z)]+s_{v2}\theta (z),\nonumber \\
	v(z)=v_1[1-\theta (z)]+v_2\theta (z),\label{ax3}
\end{eqnarray}
with
\begin{equation}
s_{c1,2}=\Delta_{1,2}+\frac{k_\perp^2}{2m_{c1,2}},\
s_{v1,2}=\Delta_{1,2}+\frac{k_\perp^2}{2m_{v1,2}}, \label{ax4}
\end{equation}
where $\Delta_{1,2}$ and $m_{c,v,1,2}$ are, respectively, the energy gaps and effective masses of interface-forming  semiconductors 1 and 2. Indices $c$ and $v$ mean, respectively, the conduction and valence bands so that $m_{1c}$ denotes the electron effective mass in the conduction band of semiconductor 1, see Refs. \cite{sd1,sd2} for details. Also, $k_\perp$ is (unimportant for the interfacial problem) the electron momentum component, parallel to the interface \cite{sd1}.

The explicit form of the Schr\"odinger equation \eqref{ax1} reads
\begin{eqnarray}
	(s_c+U-\varepsilon )\, \varphi +iv(-\Delta)^{1/2}\chi =0,\label{ax5} \\
	iv(-\Delta)^{1/2}\varphi +(-s_v+U-\varepsilon )\, \chi =0.\nonumber
\end{eqnarray}
Then we can express component $\chi$ through $\varphi$ from second equation \eqref{ax5} to obtain following eigenproblem for $\varphi$
\begin{equation}\label{ax6}
(s_c(z)+U(z))\, \varphi(z) +v(z)(-\Delta)^{1/2}\left[\frac{v(z)}{-s_v(z)+U(z)-\varepsilon }\, (-\Delta)^{1/2}\varphi(z)\right] =\varepsilon \varphi(z).
\end{equation}
We note that the eigenproblem \eqref{ax6} is indeed nonlinear as the eigenenergy $\varepsilon$ enters also the denominator. Our analysis shows that for small $\varepsilon$ the problem \eqref{ax6} is reducible to that for 1D regional fractional derivative. The best way to solve it is to expand a solution in the orthogonal set of functions with Neumann boundary conditions. The complete solution of the problem \eqref{ax6} can be obtained only numerically. Our preliminary studies of that solution show that it describes many experimentally observable salient features of the interfaces like high electronic concentration at the interface (leading, e.g. to metallic conductivity, see \cite{oh} and references therein), which appears due to disorder, described by fractional derivatives. The work in this interesting direction is underway and results will be published elsewhere.

One more interesting physical problem is the onset of a chaos in  excitons, induced by Rashba spin-orbit interaction \cite{br}. This situation is described by one more "confined" (although in two dimensions, making the solution to be much more complicated than in 1D case) problem. Namely, then one is  dealing with the quantum version of the Kepler problem, i.e.hydrogen atom, see, e.g. \cite{ash}. The fractional Hamiltonian of the 2D problem has the form
 \begin{equation}\label{ov1}
\frac{(-\Delta)^{\alpha/2}}{2m}\psi(x,y)-\frac{\psi(x,y)}{r}=E\psi(x,y),
 \end{equation}
where $r=(x^2+y^2)^{1/2}$ and we once more use atomic units. Here, the spatial confinement is due to the Coulomb potential $1/r$. Similar to the case of ordinary quantum mechanics \cite{land3,bershub}, the discreet spectrum exists only at $E<0$. It can be shown that the problem \eqref{ov1} admits the separation of angular and radial variables, leaving us with ordinary differential equation for the function $\psi(r)$. The solution of latter problem is still much more complicated then those in 1D. Our preliminary analysis shows that the function $\psi(r)$ can be obtained as the expansion of the complete set of eigenfunctions, corresponding to the "ordinary" (i.e. with conventional Laplacian in Eq. \eqref{ov1}) 2D hydrogen atom. The method is similar to that of Ref. \cite{mypre0}. Note that the problem \eqref{ov1} can be formulated in 3D with additional confinement in the potential well. Latter case corresponds to the excitons, confined to quantum wells in disordered semiconductors. One more way of problem \eqref{ov1} generalization is to take into account the fractional analog of Rashba spin-orbit interaction \cite{br}. Although the latter problem becomes very complex (for example it contains now, similar to Eq. \eqref{ax1}, spinor wave function), its solution can be obtained along the lines of Ref. \cite{mypre0}, i.e. by the expansion over properly tailoed orthonormal basis with Dirichlet boundary conditions. The discussed problems, related to fractionally quantized Kepler one, are extremely important for photovoltaic applications (see Ref. \cite{pho} for appropriate references), where possible chaotic behavior \cite{she1,she2} can disrupt the solar cell functionality. Namely, the chaotic behavior has been obtained explicitly in the form of "messy" electronic trajectories \cite{she1}. At the same time, in the quantized version only weak effects like energy levels repulsion have been revealed \cite{she2}.  It is tempting to expose the quantum chaotic features (along with direct quantum trajectories simulations) of this problem by substitution of the ordinary 2D Laplacian by the fractional one in the Schr\"odinder equation \eqref{ov1}. In this case, the levels repulsion and other features like non-Poissonian energy level statistics \cite{rei} would heavily depend on the L\'evy index $\alpha$.

\bigskip
{\bf Acknowledgment:} We would like to thank Professors K. Bogdan   and  G. Grubb  for correspondence on  censored and  reflected jump-type processes,  Yanzhi  Zhang for explanations concerning Ref. \cite{duo},  B. Dybiec for explanations concerning the stopping scenario of Refs. \cite{dybiec,dybiec1},  R. Metzler for reference suggestion and  A. Pakes  for providing  access to   his  publication on killing and resurrection of Markov processes,   \cite{pakes}.


\begin{thebibliography}{99}
\bibitem{schuss} Z. Schuss, {\it Brownian dynamics at boundaries and interfaces}, {Springer, NY, 2013}.
\bibitem{carlsaw} H. S. Carlsaw and J. C. Jaeger, {\it Conduction of heat  in solids}, (Oxford Univ. Press, London, 1959).
\bibitem{grebenkov} D. S. Grebenkov,  "NMR survey of reflected Brownian motion", Rev. Mod. Phys. {\bf 79}, 1077, (2007).
\bibitem{grebenkov1} D. S. Grebenkov, "Laplacian eigenfunctions in NMR. I. A numerical tool", Concepts Magn. Reson. A {\bf 32}, 277, (2008)
\bibitem{grebenkov2} D. S. Grebenkov, "Laplacian Eigenfunctions in NMR. II. Theoretical Advances", Concepts Magn. Reson., A {\bf 34}, 264, (2009).
\bibitem{grebenkov3} D. S. Grebenkov and B.-T. Nguyen,  "Geometrical Structure of Laplacian Eigenfunctions",  SIAM Review, {\bf 55}(4), 601, (2013).
\bibitem{bickel} T. Bickel, "A note on the confined diffusion", Physica A {\bf  377}, 24,  (2007).
\bibitem{bickel0} V. Linetsky, "On the transition densities for reflected diffusions", Adv. App. Prob. {\bf 37}, 435-460, (2005)
\bibitem{risken} H. Risken, {\it The Fokker-Planck equation}, (Springer, Berlin, 1996).
\bibitem{lorinczi} K. Kaleta and J. L\H{o}rinczi, "Transition in the decay rates of stationary distributions of L\'{e}vy motion  in an energy landscape", Phys. Rev. E {\bf 93}, 022135, (2016).
\bibitem{gitterman} M. Gitterman, "Mean first passage time for anomalous diffusion", Phys. Rev. E {\bf 62}, 6065, (2000).
\bibitem{buldyrev}  S. V. Buldyrev et al., "Average time spent by L\'{e}vy flights and walks on an interval with absorbing boundaries",
Phys. Rev. E {\bf 64}, 041108, (2001).
\bibitem{gar0} P. Garbaczewski and V. Stephanovich,  "L\'{e}vy flights in inhomogeneous environments", Physica A {\bf 389}, 4419, (2010).
\bibitem{pre11} P. Garbaczewski, W. Stephanovich, "L\'{e}vy  targeting and the principle of detailed balance",  \pre {\bf 84}, 011142 (2011)
\bibitem{gar} P. Garbaczewski, "Fractional Laplacians and L\'{e}vy flights in bounded domains", Acta Phys. Pol. B {\bf 49}(5),  921, (2018)
\bibitem{gar1} P. Garbaczewski, "Killing (absorption) versus survival in random motion",  Phys. Rev. E {\bf 96}, 032104, (2017).
\bibitem{mazzolo} A. Mazzolo, "Sweetest taboo processes", J. Stat. Mech. (2018) 073204.
\bibitem{pinsky} R. G. Pinsky, ”On the convergence of diffusion processes conditioned to remain in a bounded region for large time to limiting positive recurrent processes”, Annals of Probability, {\bf 13} (2), 363, (1985).
\bibitem{gaps} W. Bao, X, Ruan, J. Shen and C. Sheng,  "Fundamental gaps of the fractional Schr\"{o}dinger operator", arXiv:1801.06517.
\bibitem{kaleta} K. Kaleta, "Spectral gap lower bound for the one-dimensional fractional Schrödinger operator in the interval ",  Studia Math. {\bf 209}, 267, (2012).
\bibitem{frank} R. Frank, "Eigenvalue bounds for the fractional Laplacian: A review", in: G. Palatucci, T. Kuusi (Eds.), {\it Recent Developments in Nonlocal Theory},  pp. 210–235. De Gruyter, Berlin, 2018.
\bibitem{collection} H.G. Sun et al, "A new collection of real world applications of fractional calculus in science and engineering", Commun. Nonlinear Sci. Numer. Simulat. {\bf 64}, 213, (2018).
\bibitem{oliveira}  F. A. Oliveira et  al., "Anomalous diffusion: A basic
mechanism for the evolution of inhomogeneous systems", Front. Phys. 7:18, (2019),
doi: 10.3389/fphy.2019.00018
\bibitem{negrete} D. del-Castillo-Negrete, "Fractional diffusion models of nonlocal transport",  Physics of Plasmas, {\bf 13}, 0822308, (2006).
\bibitem{laskin0} N. Laskin, "Fractional Schr\"{o}dinger equation", Phys. Rev. E {\bf 66}, 056108, (2002).
\bibitem{laskin1}  N. Laskin, "Fractional quantum mechanics and L\'{e}vy path integrals",
 Phys. Lett. A {\bf  268},   298, (2000).
\bibitem{laskin}  N. Laskin, {\it Fractional quantum mechanics}, (World Scientific, Singapore, 2018).
\bibitem{bucur} C. Bucur and E. Valdinoci,  {\it Nonlocal diffusion and applications},  Lecture Notes of the Unione Matematica Italiana, vol. 20, (Springer International Publishing   Switzerland,  2016).
\bibitem{vazquez} J. L. V\'{a}zquez, "The Mathematical Theories of Diffusion: Nonlinear and Fractional Diffusion", in: M. Bonforte, G. Grillo (Eds.), {\it  Nonlocal and Nonlinear Diffusions and Interactions: New Methods and Directions},  Lecture Notes in Mathematics vol.
2186, (Springer, NY, 2017)
\bibitem{abtangelo}  N. Abatangelo and E. Valdinoci,  "Getting acquainted with the fractional Laplacian", arXiv:1710.1156.
\bibitem{what} A. Lischke et al, "What is the  fractional Laplacian", arXiv:1801.09767.
\bibitem{servadei} R. Servadei and E. Valdinoci, "On the spectrum of two different  fractional operators", Proceedings of the Royal Society of Edinburgh Section A: Mathematics, {\bf 144}(4), 831, (2014).
\bibitem{dybiec} B. Dybiec, E. Gudowska-Nowak and P. H\"{a}nggi, "L\'{e}vy-Brownian motion on finite intervals: Mean-first passage  time analysis", Phys. Rev. E {\bf 73}, 046104, (2006).
\bibitem{dybiec1}  B. Dybiec, E. Gudowska-Nowak, E. Barkai and A. A. Dubkov, "L\'{e}vy flights  versus L\'{e}vy walks in   bounded domains", Phys. Rev. E {\bf 95}, 052102, (2017).
\bibitem{dubkov}A. A. Kharcheva et al, "Spectral characteristics of steady-state L\'{e}vy flights in confinement potential
profiles", J. Stat. Mech. vol.  2016 (5),054039.
\bibitem{broeck} V. Balakrishnan, C. Van den Broeck and P. H\"{a}nggi, "First-passage times of non-Markovian processes: The case of reflecting boundary", Phys. Rev. {\bf 38}, (8), 4213, (1988).
\bibitem{denisov} S. I. Denisov, W. Horsthemke and P. H\"{a}nggi, "Steady-state       L\'{e}vy flights in a confined domain", Phys. Rev. E {\bf 77}, 061112, (2008).
\bibitem{bogdan} K. Bogdan, K. Burdzy and Z.-Q. Chen, "Censored stable processes",   Probab. Theory Relat. Fields, {\bf 127}, 89, (2002).
\bibitem{refl} Q.-Y. Guan \& Z.-M. Ma, "Reflected symmetric $\alpha $-stable processes and regional fractional Laplacian", Probab. Theory Relat. Fields {\bf 134}, 649 (2006).
\bibitem{guan} Q.-Y. Guan \& Z.-M. Ma, "Boundary problems for fractional Laplacians", Stochastics and Dynamics, {\bf 5}(3), 385, (2005)
\bibitem{warma} M. Warma, " A fractional Dirichlet-to-Neumann operator on bounded Lipschitz domains", Commun. Pure Appl. Analysis, {\bf 14}(5), 2043, (2015).
\bibitem{barles} B. Barles, E. Chasseigne, C. Georgelin and E. R. Jacobsen, "On Neumann problems for non-local equations set in a half-space", Trans. Amer. Math. Soc. {\bf 366}, 4873, (2014).
\bibitem{dipierro}  S. Dipierro, X. Ris-Oton and E. Valdinoci, "Nonlocal problems with Neumann boundary conditions", Revista Matematica Iberoamericana 33(2), 377, (2017).
\bibitem{abtangelo1} N. Abtangelo, "A remark on nonlocal Neumann conditions for the fractional Laplacian", arXiv:1712.00320,  (2018).
\bibitem{grubb}  G. Grubb, "Fractional-order operators : Boundary problems, heat equations", arXiv:1712.01196, to appear in  Springer Proceedings in Mathematics and Statistics: "New Perspectives in Mathematical Analysis - Plenary Lectures, ISAAC 2017, Vaxjo Sweden".
\bibitem{zaba} P. Garbaczewski, M. Zaba,  "Nonlocal random motions and the trapping problem",  Acta Phys. Pol. B 46, 231 (2015).
\bibitem{wada} A. H. O. Wada and T. Vojta, "Fractional Brownian motion with a  reflecting wall". Phys. Rev. E {\bf 97}, 020102, (2018).
\bibitem{neel}  M.-C. N\'{e}el, L. Di Pietro and N. Krepysheva, "Enhanced diffusion in a bounded domain", IFAC Proceedings Volumes, {\bf 39}(11), 448, (2006).
\bibitem{metzler} A. Compte, R. Metzler and J. Camacho, "Biased continuous time random walks between parallel plates", Phys. Rev. E {\bf 56}(2), 1445, (1996).
\bibitem{metzler1} R. Metzler and J. Klafter, "Boundary value problems for
fractional diffusion equations", Physica A {\bf 278}, 107, (2000).
\bibitem{meer}  B. Beumer, M. Kov\'{a}cs, M. M. Meerschaert and H. Sankaranarayanan, "Reprint of: Boundary conditions for fractional diffusion", J. Comput. Applied Math., {\bf 339}, 414, (2018).
\bibitem{kinley} S. A. McKinley and H. Nguyen,"Anomalous diffusion and the Generalized Langevin Equation", SIAM J. Math. Analysis {\bf 50} (5), 5119,
2017).
\bibitem{vainstein} M. H. Vainstein et al., "Non-exponential relaxation for anomalous diffusion", Europhys Lett. {\bf 73} (5), 726, (2006).
\bibitem{metzler2}  T. Guggenberger et al.,  " Fractional Brownian motion in a finite interval:Correlations effect,  depletion or accretion zones of
particles near boundaries,  New Journal of Physics, {\bf 21}, 022002, (2019).
\bibitem{kwasnicki}  M. Kwa\'{s}nicki, "Ten equivalent definitions of the fractional Laplace operator", Fractional Calculus \& Applied Analysis, {\bf 20}(1), 7, (2017).
\bibitem{kwa} M. Kwa\'{s}nicki, "Fractional Laplace operator and its properties", in: {\it  "Handbook of fractional calculus with applications"},  vol. 1, (Eds.) A. Kochuba and Y. Luchko, de Gruyter, Berlin,  2019,  to be published.
\bibitem{vondr} R. Song and Z. Vondracek, "Potential theory of subordinate killed Brownian motion in a domain", Probab. Theory. Relat. Fields, {\bf 125}, 578, (2003).
\bibitem{kulczycki} T. Kulczycki, "Eigenvalues and Eigenfunctions for Stable Processes", chap. 4, pp. 73-86,  in: K. Graczyk and A. Stos (Eds.), "Potential Analysis of Stable
Processes and its Extensions", LNM vol. 1980, (Springer-Verlag, Berlin, 2009).
\bibitem{duo} S. Duo, H. Wang and  Y. Zhang, "A comparative study of nonlocal diffusion operators related to the fractional Laplacian",     Discrete \& Continuous Dynamical Systems - B {\bf 22}(11),  1, (2018).
\bibitem{duo1}  S. Duo and Y. Zhang,  "Computing the ground and first excited states of the fractional Schr\"{o}odinger equation in an ininite potential well",  Commun. Comput. Phys., {\bf 18}, 321-350, (2015).
\bibitem{zaba0} M. \.{Z}aba and P. Garbaczewski, "Solving fractional Schrödinger-type spectral problems: Cauchy oscillator and Cauchy well", J. Math. Phys. {\bf 55},  092103, (2014).
\bibitem{zaba1} M. \.{Z}aba and P. Garbaczewski, "Nonlocally induced (fractional) bound states: Shape analysis  in the infinite Cauchy well", J. Math. Phys. {\bf 56}, 123502, (2015).
\bibitem{mypre0}  E. V. Kirichenko, P. Garbaczewski, V. Stephanovich and M. \.{Z}aba, "L\'{e}vy flights in an infnite potential well as a hypersingular Fredholm problem", Phys. Rev. E {\bf 93}, 052110,  (2016).
\bibitem{mypre} E. V. Kirichenko, P. Garbaczewski, V. Stephanovich and M. \.{Z}aba, "Ultrarelativistic (Cauchy) spectral problem in the infinite well", Acta Phys. Pol. B {\bf 47}(5), 1273-1291, (2016).
\bibitem{teso}  N. Cusimano, F. del Teso, L. Gerardo-Giorda and G. Pagnini,  "Discretizations of the Spectral Fractional Laplacian on General Domains with Dirichlet, Neumann, and Robin Boundary Conditions", SIAM j. Numer. Anal., {\bf 56}(3), 1243, (2018).
\bibitem{musina} R. Musina and A. I. Nazarov, "On fractional Laplacians", Comm. Partial. Diff. Eqs. {\bf 39}(9), 1780, (2014).
\bibitem{vondr1} P. Kim, R. Song and Z. Vondracek, "On the boundary theory of subordinate killed L\'{e}vy processes", arXiv: 1705.02595, (2017).
\bibitem{luchko}  Y. Luchko, "Fractional Schr\"{o}dinger equation for a particle moving in a potential well", J. Math. Phys., {\bf 54}, 012111, (2013).
\bibitem{bayin}  S. S. Bayin, "Consistency problem of the solution of the space fractional Schr\"{o}dinger equation", J. Math. Phys., {\bf 54},  092101, (2013).
\bibitem{wei} Y. Wei, "The infinite square well problem in standard, fractional and relativistic quantum mechanics", Int. J. Theor. Math. Phys, {\bf 5}(4), 58, (2015)
\bibitem{kaleta1}  K. Kaleta, M. Kwa\'{s}nicki and J. Ma{\l}ecki, "Asymptotic
estimate of eigenvalues of pseudo-differential operators in an interval", J. Math Anal. Appl., {\bf 439}, 896, (2016).
\bibitem{stos} T. Kulczycki, M. Kwa\'{s}nicki, J. Ma{\l}ecki  and A. Stos, "Spectral properties of the Cauchy process on half‐line and interval", Proc. London Math. Soc., {\bf 101}(2), 589, (2010).
\bibitem{kwasnicki1} M. Kwa\'{s}nicki, "Eigenvalues of the fractional Laplace operator in the interval", J. Funct. Anal., {\bf 262}(5), 2379, (2012).
\bibitem{malecki} J. L\H{o}rinczi and J. Ma{\l}ecki, "Spectral properties of the massless relativistic harmonic oscillator", J. Diff. Equations, {\bf 253},2846, (2012).
\bibitem{durugo} S. Durugo and J. L\H{o}rinczi, "Spectral properties of the massless relativistic quartic oscillator", J. Diff. Equations,  {\bf 264}(5), 3775, (2018).
\bibitem{pakes} A. G. Pakes, "Killing and resurrection of Markov processes", Commun. Statist. Stochastic models, {\bf 13}(2), 255, (1997).
\bibitem{evans}  M. R. Evans and S. N. Majumdar, "Diffusion with stochastic resetting", Phys. Rev. Lett. {\bf 106}, 160601, (2011).
\bibitem{evans1}  M. R. Evans and S. N. Majumdar, "Diffusion with stochastic resetting  in arbitrary spatial dimension", J. Phys. A: Math. Theor., {\bf 47}, 285001, (2014).
\bibitem{durang} X. Durang, M. Henkel and H. Park, "The statistical mechanics of the
coagulation - diffusion process with a stochastic reset", J. Phys. A: Math. Theor. {\bf  47}, 045002, (2014).
\bibitem{dual} P. Garbaczewski, "Modular Schr\"{o}dinger equation and dynamical duality", Phys. Rev. E {\bf 78}, 031101, (2008).
\bibitem{kulczycki1} T. Kulczycki, "Intrinsic ultracontractivity for symmetric stable processes",  Bull. Polish  Acad. Sci. Math. 46,  325, (1998).
\bibitem{grzywny} T. Grzywny, "Intrinsic Ultracontractivity for L\'{e}vy Processes", Prob. Math. Statistics, {\bf 28}(1), 91, (2008).
\bibitem{chen} X. Chen and J. Wang, "Intrinsic ultracontractivity for general
 L\'{e}vy processes on bounded open sets", Illinois J. Math. {\bf 58}(4), 1117, (2014).
 \bibitem{banuelos} R. Ba\~{n}uelos, T. Kulczycki and   P. J. Mendez-Hernandez, "On the shape of the ground state eigenfunction for stable processes",  Potential Analysis, {\bf 24}, 205, (2006).
 \bibitem{guide} E. Di Nezza, G. Palatucci and E. Valdinoci, "Hitchhiker’s  guide to the fractional Sobolev spaces", Bull. Sci Math. {\bf 136}(5), 521, (2012).
 \bibitem{duo2}   S. Duo, H. W. van Wyk, and Yanzhi Zhang, "A novel and accurate finite difference method for the fractional Laplacian and the fractional Poisson problem", J. Comput. Phys., {\bf 355}, 233, (2018).
\bibitem{pezzo}  J. P. Borthagaray, L. M. Del Pezzo and S. Martinez, "Finite element approximation for the fractional eigenvalue problem", J. Sci. Comput. {\bf 77}, 308, (2018).
\bibitem{diaz} J. I. Diaz, D. G\'{o}mez-Castro and J. L. V\'{a}zquez,  "The fractional Schr\"{o}dinger equation with general nonnegative potentials. The weighted space approach", arXiv: 1804.08398, (2018).
\bibitem{abr} Handbook of Special Functions. Ed. M. Abramowitz, I.A. Stegun.
National Bureau of Standards, NY, 1964.
\bibitem{dybiec2} B. Dybiec, private communication.
\bibitem{dyda}  B. Dyda, "Fractional calculus for power functions", Fract. Calc. Appl. Anal. {\bf 15} (4), 536,  (2012).
\bibitem{private} M. \.{Z}aba, private communication.
\bibitem{land3} L. D. Landau and E. M. Lifshits, {\em{Quantum Mechanics. Nonrelativistic Theory}} (Pergamon Press, Oxford, 1995).
\bibitem{stef} P. Garbaczewski and V. A. Stephanovich, "L\'{e}vy flights nd nonlocal quantum  dynamics", J. Math. Phys. {\bf 54}, 072103, (2013).
\bibitem{bershub} F.A. Berezin and M.A. Shubin  {\em{The Schr\"{o}dinger Equation}} (Kluwer, Dordrecht, 1991).
\bibitem{zu} I. Zuti{\v c}, J. Fabian, S. Das Sarma, "Spintronics: Fundamentals and applications", \rmp \ {\bf 76}, 323 (2004)
\bibitem{gvi} A. H. Castro Neto, F. Guinea, N. M. R. Peres, K. S. Novoselov, and A. K. Geim, "The electronic properties of graphene", \rmp \ {\bf 81}, 109 (2009).
\bibitem{sar}  S. Das Sarma, S. Adam, E. H. Hwang, and E. Rossi, "Electronic transport in two-dimensional graphene",  \rmp \ {\bf 83}, 407 (2011).
\bibitem{oh} A. Ohtomo and H. Y. Hwang, "A high-mobility electron gas at the LaAlO$_3$/SrTiO$_3$ heterointerface.", Nature {\bf 427}, 423 (2004).
\bibitem{sd1} V. A. Stephanovich, V. K. Dugaev and J. Barnas, "Two-dimensional electron gas at the
 LaAlO$_3$/SrTiO$_3$ inteface with a potential barrier", Phys. Chem. Chem. Phys. {\bf 18}, 2104 (2016).
\bibitem{sd2} V. A. Stephanovich and V. K. Dugaev, "Macroscopic description of the two-dimensional LaAlO$_3$/SrTiO$_3$ interface", \prb \ {\bf 93}, 045302 (2016).
\bibitem{br} Yu. A. Bychkov and E. I. Rashba, "Properties of a 2D electron gas with lifted spectral degeneracy", JETP Lett., {\bf 39}, 78 (1984).
\bibitem{ash} N.W. Ashkroft and N. D. Mermin, {\em{Solid State Physics}} (Harcourt, New York, 1976).
\bibitem{pho} W. S. Yang, J. H. Noh, N. J. Jeon, Y. C. Kim, S. Ryu, J. Seo and
S. I. Seok, "High-performance photovoltaic perovskite layers fabricated through intramolecular exchange", Science, {\bf 348}, 1234 (2015).
\bibitem{she1} V.A. Stephanovich and E. Ya. Sherman, "Chaotization of internal motion of excitons
in ultrathin layers by spin–orbit coupling", Phys. Chem. Chem. Phys. {\bf 20}, 7836 (2018).
\bibitem{she2}  V.A. Stephanovich, E. Ya. Sherman, N.T. Zinner and O.V. Marchukov, "Energy-level repulsion by spin-orbit coupling in two-dimensional Rydberg excitons", \prb \ {\bf 97}, 205407 (2018).
\bibitem{rei} L. E. Reichl, {\em{The Transition to Chaos. Conservative Classical
Systems and Quantum Manifestations}} (Springer-Verlag, New York, 2nd edn, 2004).
\end{thebibliography}
\end{document}